\documentclass[twocolumn,english,aps,floatfix,showpacs,prb]{revtex4-1}
\usepackage[latin9]{inputenc}
\usepackage{amsmath}
\usepackage{amssymb}
\usepackage{graphicx}

\makeatletter
\usepackage{color}
\usepackage{textcomp}

\usepackage{dcolumn}
\usepackage{bm}
\usepackage{ulem}
\usepackage[english]{babel}
\@ifundefined{textcolor}{}{
 \definecolor{BLACK}{gray}{0}
 \definecolor{WHITE}{gray}{1}
 \definecolor{RED}{rgb}{1,0,0} 
 \definecolor{GREEN}{rgb}{0,1,0}
 \definecolor{BLUE}{rgb}{0,0,1}
 \definecolor{CYAN}{cmyk}{1,0,0,0}
 \definecolor{MAGENTA}{cmyk}{0,1,0,0}
 \definecolor{YELLOW}{cmyk}{0,0,1,0}
 }

\usepackage{babel}

\makeatother

\usepackage{babel}
\begin{document}

\title{Variational Monte Carlo method for fermionic models combined with
tensor networks and applications to the hole-doped two-dimensional Hubbard model }
\author{Hui-Hai Zhao$^{1,2}$ }
\email{zhaohuihai@solis.t.u-tokyo.ac.jp}

\author{Kota Ido$^{1}$}

\author{Satoshi Morita$^{2}$}

\author{Masatoshi Imada$^{1}$}
\email{imada@ap.t.u-tokyo.ac.jp}

\affiliation{$^1$Department of Applied Physics, The University of Tokyo, Hongo, Bunkyo-ku,
Tokyo 113-8656, Japan}
\affiliation{$^2$Institute for Solid State Physics, The University of Tokyo, Kashiwanoha,
Kashiwa, Chiba 277-8581, Japan}

\selectlanguage{english}%

\date{\today}

\begin{abstract}
The conventional tensor-network states employ real-space product states
as reference wave functions. Here, we propose a many-variable variational
Monte Carlo (mVMC) method combined with tensor networks
by taking advantages of both to study fermionic models. The variational
wave function is composed of a pair product wave function operated
by real space correlation factors and tensor networks. Moreover, we
can apply quantum number projections, such as spin, momentum and lattice
symmetry projections, to recover the symmetry of the wave function
to further improve the accuracy. We benchmark our method for one-
and two-dimensional Hubbard models, which show significant improvement
over the results obtained individually either by mVMC or by tensor
network. We have applied the present method to hole doped Hubbard
model on the square lattice, which indicates the stripe charge/spin
order coexisting with a weak $d$-wave superconducting order in the ground state for the
doping concentration less than 0.3, where the stripe oscillation period gets longer with increasing hole concentration. 
The charge homogeneous and highly superconducting state also exists as a metastable excited state for the doping concentration less than 0.25. 
\end{abstract}


\maketitle
\section{\label{sec:intro}Introduction}

Finding the ground state of strongly correlated electron systems is
one of most challenging problems in condensed matter physics. Since
exact solutions only exist in few systems, deeper understanding of
ground state properties strongly relies on efficient and accurate
numerical algorithms. For example, one can employ exact diagonalization
(ED) to find the wave function accurately, but it is only applicable
for very small size systems. The density matrix renormalization group
(DMRG)\citep{DMRG1992} is very efficient and accurate for one dimensional
systems, but it becomes inefficient for two and higher dimensional
systems. The quantum Monte Carlo methods\citep{QMC2001} suffer from
the sign problem in general fermionic and geometrically frustrated
systems.

In the past years, the tensor network algorithms have been widely
developed\citep{Niggemann1997,PTP2001Nishino,PEPS2004,MERA,TRG2007,PRL2007Vidal,SimpleUpdate2008,TEFR,SRG2009,SingleLayer2011,HOTRG2012,PESS2014,TNR},
which are shown to be promising numerical tools. One of the simplest
tensor network state is the matrix product state (MPS), which is the
variational wave function of the DMRG method\citep{Ostlund1995,DMRGpbc2004}.
A natural generalization of the MPS to two dimensions is the projected
entangled pair state (PEPS)\citep{PEPS2004}, which satisfies the
area law of entanglement entropy\citep{PEPS2004}. Besides PEPS, various
types of tensor network states have been introduced, such as multi-scale
entanglement renormalization ansatz (MERA)\citep{MERA}, tree tensor
network states\citep{TTN2009} and projected entangled simplex states\citep{PESS2014}.
These tensor network wave functions are usually expressed in a real
space basis which may become inefficient to capture the large amount
of entanglement in itinerant fermionic systems. For example, the free
fermion model, which can be exactly represented as a product state
in momentum space, is extremely difficult to accurately treat by infinite
fermionic PEPS algorithm\citep{fPEPS2010}.

The variational Monte Carlo (VMC)\citep{Ceperley1977} method can
be applied to study relatively large system sizes, and there is no
sign problem in studies of fermionic and frustrated systems. However,
the result is subject to be biased depending on the form of variational
wave functions. In the region where various competing phases have
very closed energy, it is very challenging to determine the correct
ground state.

Sorella has developed stochastic reconfiguration (SR)\citep{SR2001}
method to stably optimize large number of parameters, which makes
it possible to extend the variational wave function to substantially
alleviate the bias. With the combination of pair-product wavefunctions
and the correlation factors such as Gutzwiller~\cite{Gutzwiller1963},
and Jastrow~\cite{Jastrow1955} factors, as well as the quantum number
projections, thousands and ten thousands of the variational parameters
have been optimized, which has enabled accurate estimates of the competing
ground states in terms of the comparisons with available exact results~\cite{Edegger2007,mVMC2008,Becca2011,Misawa2014PRB}.
The applications have achieved fruitful and reproducible comparisons
with the experimental results, for instance for the iron-based superconductors,
if the method is applied to the \textit{ab initio} effective Hamiltonians\cite{Misawa2014NatComm}.
However, how to further systematically remove the bias in the variational
wave functions and reach better accuracy is still left as a challenge.

The Monte Carlo sampling techniques have been proposed as a prescription
to reduce the computational cost of tensor network contraction in
the MPS\citep{MPSVMC2007}, PEPS\citep{PEPSVMC2011} and MERA\citep{MERAVMC2012}.
Another advantage of employing the Monte Carlo sampling into tensor
network methods is that it is possible to choose various types of
suitable reference basis beyond the real space basis. In the VMC study
of correlator product states\citep{CPS2009}, a Pfaffian pairing wave
function has been used\citep{Neuscamman2012}. In the study of one
dimensional fermionic system, the free fermion Slater determinant
has been used as the reference wave function of the MPS\citep{MPPSVMC2012},
which achieves higher accuracy than the conventional MPS method with
the same bond dimension. Ref. \onlinecite{TPPSVMC2015} has generalized
this idea to two dimensional systems, which employs the PEPS with
various kinds of reference wave functions, such as Jastrow-type, free
fermion, $d$-wave BCS and spin density wave states, in order to choose
a suitable reference wave function that captures the key features
of the systems.

In this paper, we employ a combination of tensor network and reference
wave function which consists of a Pfaffian pairing wave function and
real space correlation factors, 
and the tensor networks can be regarded as diagonal correlation projectors, which act in the same way as that in
Refs. \onlinecite{MPPSVMC2012} and \onlinecite{TPPSVMC2015}. 
In order to provide more flexible
representation, we optimize all the variational parameters both in
the part of the VMC and the tensor network simultaneously. Moreover,
we can apply quantum number projections, such as those to restore
the spin, momentum and lattice symmetries to further improve the accuracy.
As a result, highly accurate ground state wave functions are obtained
within a computationally tractable size of tensor bond dimension beyond
the accuracy of each method if applied separately. Moreover, the accuracy
can be continuously improved with the increase of the tensor bond
dimension, thus providing a systematic way of removing the bias in
the VMC. The combination with the VMC is particularly powerful if
the tensor network method suffers from the entanglement entropy remaining
beyond the area law as in the case of the itinerant fermion systems.

In the latter part of this paper, we show a fruitful application to 
a strongly correlated system: Hubbard model on the square lattice. 
Although the relevance of the Hubbard model for the mechanism of the high $T_{\rm c}$ superconductivity is an open issue, the ground states of the hole-doped two-dimensional Hubbard model
has been extensively studied and debated for decades, because it
is a simplest model of the cuprate superconductors and it captures 
several important experimental aspects. However, it was also suggested that strong 
competitions exist among different states including $d$-wave superconductivity 
and various charge inhomogeneous states
such as stripe order and phase separation.
Accurate
determination of the phase diagram in the plane of the carrier density
and the electron correlation strength is still an open question. The
present method opens a possibility of studying the model at the highest
accuracy among that ever studied. We show the numerical results of
the $d$-wave superconducting correlations and various spin and charge
correlations that indeed reveal the severe competition of the these two types
of orders and clarifies how they are compromised in the best estimates
of the ground states.

This paper is organized as follows. In Sec. \ref{sec:method}, we
introduce our variational wave function, and describe how to optimize
large number of parameters in variational Monte Carlo methods. In
Sec. \ref{sec:benchmark}, we present our benchmark results for the
Hubbard model in one and two dimensions. In  Sec.\ref{dopedHubbard}, we show results obtained
by applying the present method to the hole doped Hubbard model to
understand the interplay between the charge/stripe order and the $d$-wave
superconductivity. Finally, we summarize the methods and results with
the future scope in Sec. \ref{sec:summary}.

\section{\label{sec:method}Numerical methods}
\subsection{Model}

Although the method presented here is applicable to general fermionic
systems on lattices, to make a presentation clearly understandable,
we keep in mind the Hubbard model with the hopping amplitude $t$
between the nearest neighbor sites $\left\langle i,j\right\rangle$
and the onsite Coulomb repulsion $U$ defined by 
\begin{equation}
H=-t\sum_{\left\langle i,j\right\rangle ,\sigma}\left(c_{i\sigma}^{\dagger}c_{j\sigma}+h.c.\right)+U\sum_{i}n_{i\uparrow}n_{i\downarrow},\label{Hubbard}
\end{equation}
where $c_{i\sigma}^{\dagger}$ and $c_{j\sigma}$ are the creation
and annihilation operators of electron with spin $\sigma$ at the
$i$-th and $j$-th site, respectively, and $n_{i\sigma}=c_{i\sigma}^{\dagger}c_{i\sigma}$
is the number operator. Here, we take the energy unit $t=1$. We mainly
consider the model on the $L\times L$ square lattice.

\subsection{Variational wave function ansatz}

The purpose of our work is to provide a flexible variational wave
function which can be applied to efficiently capture the key features
of the systems. We apply the wave function employed in the mVMC\citep{mVMC2008}
as the reference wave function of tensor network states, which is
expressed as,

\begin{equation}
\left|\phi_{{\rm ref}}\right\rangle =\mathcal{P}\mathcal{L}^{{\rm Space}}\mathcal{L}^{S}\mathcal{L}^{K}\left|\phi_{{\rm pair}}\right\rangle ,\label{eq:mVMC}
\end{equation}
where $\mathcal{P}$ is the product of real-space correlation factors
\begin{equation}
\mathcal{P}=\mathcal{P}_{{\rm G}}\mathcal{P}_{{\rm J}}\mathcal{P}_{{\rm d-h}}^{{\rm ex}},\label{eq:correlationFactor}
\end{equation}
in which $\mathcal{P}_{{\rm G}}$ is the Gutzwiller factor\citep{Gutzwiller1963}
that punishes (enhances) the double occupation of electrons on the
same site defined as

\begin{equation}
\mathcal{P}_{{\rm G}}=\exp\left(-\sum_{i}g_{i}n_{i\uparrow}n_{i\downarrow}\right)
\end{equation}
to take into account the local correlation effects, $\mathcal{P}_{J}$
is the Jastrow factor\citep{Jastrow1955} which accounts for long-ranged
density correlations through two-body operators defined as

\begin{equation}
\mathcal{P}_{{\rm J}}=\exp\left[\frac{1}{2}\sum_{i\neq j}v_{ij}\left(n_{i\uparrow}+n_{i\downarrow}\right)\left(n_{j\uparrow}+n_{j\downarrow}\right)\right],
\end{equation}
and $\mathcal{P}_{d-h}^{{\rm ex}}$ is the doublon-holon correlation
factor\citep{DH1982,DH1990} 
\begin{equation}
\mathcal{P}_{{\rm d-h}}^{{\rm ex}}=\exp\left(-\sum_{m=0}^{4}\sum_{l}\alpha_{\left(m\right)}^{\left(l\right)}\sum_{i}\xi_{i\left(m\right)}^{\left(l\right)}\right),
\end{equation}
where $\alpha_{\left(m\right)}^{\left(l\right)}$ are variational
parameters, and $\xi_{i\left(m\right)}^{\left(l\right)}=1$ when a
doublon (holon) exists at the site $i$ and $m$ holons (doublons)
surround at the $l$-th nearest neighbor sites and otherwise, $\xi_{i\left(m\right)}^{\left(l\right)}=0$.
The range of $l$ is taken short, normally up to the second nearest
neighbors.

The two-body part $\phi_{{\rm pair}}$ is expressed in the real space
representation
\begin{equation}
\left|\phi_{{\rm pair}}\right\rangle =\left(\sum_{i,j=1}^{N_{s}}\sum_{\sigma\sigma^{\prime}}f_{i\sigma j\sigma^{\prime}}c_{i\sigma}^{\dagger}c_{j\sigma^{\prime}}^{\dagger}\right)^{N_{e}/2}\left|0\right\rangle ,\label{eq:pairProd}
\end{equation}
where $f_{i\sigma j\sigma^{\prime}}$ denotes the variational parameters,
and $N_{s}$, $N_{e}$ are the number of sites and electrons, respectively.
Although the spins $\sigma$ and $\sigma^{\prime}$ are taken to be
opposite in this paper to impose the constraint of the singlet pair,
it can be easily generalized to arbitrary spins for better accuracy.
The number of variational parameters are at most $N_{s}$, $N_{s}^{2}$,
and $N_{s}^{2}$ for $g_{i},v_{ij}$ and $f_{ij}$, respectively.
To save the computational cost, these numbers are sometimes reduced
by imposing the sublattice translational invariance. In the case of
the sublattice size $N_{{\rm sub}}$, the numbers are reduced to $N_{s}N_{{\rm sub}}$
for $v_{ij}$ and $f_{ij}$, while $g_{i}$ is taken site independent
in the following study.

The two-body part $\phi_{{\rm pair}}$ may contain symmetry broken
phases for better accuracy. On top of it, the space group, spin and
momentum quantum number projection operators $\mathcal{L}^{{\rm Space}}$,
$\mathcal{L}^{S}$ and $\mathcal{L}^{K}$ can be employed to recover
the space group, spin SU(2) symmetry and lattice translational symmetry
of the wave function to further improve the accuracy, because these
symmetries are preserved for finite-size systems in most of the Hamiltonians
we study. In this paper, we focus on the case of the singlet $S=0$,
and the total momentum zero $K=0$ because they are satisfied in most
of the ground state of models. We impose the space group symmetry
later. We also note that by imposing these projections, one can also
study the lowest excited state of the specified quantum number.

Combining with the tensor network algorithm, the variational wave
function we are going to optimize is 
\begin{eqnarray}
\left|\Psi\right\rangle  & = & \sum_{q_{1},q_{2},\ldots,q_{N_{s}}=1}^{d}\mathcal{P}\left(\mathcal{L}^{K=0}\mathcal{L}^{C_{4}}\mathcal{M}\right)\left|q_{1},q_{2},\ldots,q_{N_{s}}\right\rangle \nonumber \\
 &  & \times\left\langle q_{1},q_{2},\ldots,q_{N_{s}}\right|\mathcal{L}^{S=0}\mathcal{L}^{K=0}\left|\phi_{{\rm pair}}\right\rangle, 
\label{eq:TNVMC}
\end{eqnarray}
where $\mathcal{M}$ is the tensor network, and the physical index
$q_{m}=1,2,\ldots,d$ represents $d$ local states at site $m$.
Since the tensor network
may break the lattice translational and rotational symmetry explicitly,
the momentum projection $\mathcal{L}^{K=0}$ and space-group symmetry
projection (for instance, the $\mathcal{L}^{C_{4}}$ to restore the
$C_{4}$ rotational symmetry in the case of the square lattice) improve
the wave function by recovering the symmetries. We apply the same
quantum number projections to the Pfaffian pair state, because this
preconditioning further improve the state.

The remaining task is to choose the appropriate tensor network $\mathcal{M}$
in Eq. (\ref{eq:TNVMC}). In one dimension, it is natural to choose
$\mathcal{M}$ as an MPS
\begin{eqnarray}
 &  & \mathcal{M}\left|q_{1},q_{2},\ldots,q_{N_{s}}\right\rangle \nonumber \\
 & = & {\rm Tr}\left(A^{1}\left[q_{1}\right]A^{2}\left[q_{2}\right]\cdots A^{N_{s}}\left[q_{N_{s}}\right]\right)
\left|q_{1},q_{2},\ldots,q_{N_{s}}\right\rangle,
\end{eqnarray}
because it holds the lattice translational symmetry and the computational
cost is as low as $O\left(D^{3}\right)$ for the periodic boundary
MPS, where $A^{m}[q_{m}]$ are $D\times D$ matrices. 
The operation Tr is to trace out all the matrix indices. 

In two dimensional systems, various types of tensor network states
may be employed. Among them we select and employ to meet several requirements.
The first requirement is that the network structure should keep the
lattice symmetry as much as possible. Therefore, the MPS will not
be considered. The second requirement is that the tensor network exactly
reproduces the ground state in the limit of infinite bond dimensions with nonzero reference basis,
so that the accuracy is improved systematically. Therefore, the string
bond state\citep{SBS2008} is ruled out, because it does not guarantee
the covering of the whole Hilbert space even when the infinite bond
dimensions are taken. The third is that we prefer a tensor network
in which the contraction can be done without truncation, so that the
variational principle will not be broken during the optimization.
Therefore, the PEPS is not employed, since the computational cost
grows exponentially if the truncation is excluded.

In order to satisfy the above requirements, we build up the tensor
network based on the idea of tree tensor network (TTN) as described
in Appendix \ref{Apdx:TTN}. The TTN is a two dimensional tensor network
which can be contracted exactly in polynomial time. If we employ the
Monte Carlo sampling on the real space configuration, the physical
indices of the leaf tensors are fixed, so every leaf tensor becomes
a vector. Therefore, we can start from the contraction of the vector
at the leaf tensor and then continue the contraction of rank 3 tensors
at the higher hierarchical levels, of which the computational cost
scales as $O\left(N_{s}D^{3}\right)$, where $D$ is the dimension of
the virtual indices.

In the TTN for the $L\times L$ lattice, the number of bonds of tensors
connecting any 2 sites is at most $O\left(\log\left(L\right)\right)$,
which has a potentiality to capture the long range entanglement efficiently.
However, the number of bonds of tensors connecting 2 neighboring sites
can be as large as $O\left(\log\left(L\right)\right)$, if the neighboring
sites belong to a different large block, which poses a limitation.
For instance some of nearby sites belonging to different blocks are
sparsely connected via internal-node tensors at a high hierarchical
level.

\begin{figure}
\includegraphics[width=8cm]{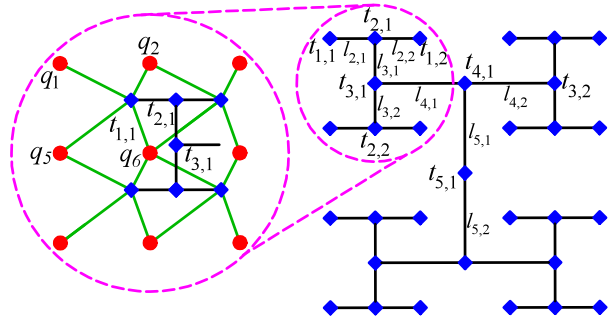} \caption{(color online) 
Example of an FTTN for a $4\times4$ lattice. Red solid
circles represent the lattice sites and blue diamonds represent the
tensors in the FTTN. $t_{1,i}$ is the leaf tensor which contains 4
physical lattice points bridged by tensor indices represented by green
bonds in the left dashed circle, and 
$t_{2,i}$, $t_{3,i}$, $t_{4,i}$, $\cdots$
are internal-node tensors which are connected only by
virtual indices represented as black bonds. Every 2 neighboring leaf
tensors share 2 common physical indices. Note that the two sites (such
as the two right bottom solid red circles (sites)) at the border of
a block (such as the dashed red circle) is connected also through
a leaf tensor in a nearby block (such as the block below the red circle.)
Therefore, these two sites (red circles) are shared and connected
by green-bond tensor indices (not shown) with the block below them.
The most right bottom site is shared by the block in the right to
the red dashed circle as well. }
\label{fig:FTTN} 
\end{figure}

We modify the standard network structure of the TTN to improve the
efficiency while keeping the computational cost in the same order.
The aim is to keep the neighboring sites as neighbors also in the
tensor network. For this purpose, we propose a fat tree tensor network
(FTTN), which contains redundant physical indices in the leaf node
tensors. Consider an $L\times L$ square lattice system in which the
local Hilbert space dimension of each site is $d$. As shown in Fig.
\ref{fig:FTTN} for $4\times4$ lattice, the FTTN is composed of a
set of tensors $t_{j,i}$, 
where $j=1,2,\cdots,R+1\left(R=\log_{2}N_s\right)$, and
$i=1,2,\ldots,N_{s}/2^{j-1}$. The FTTN is connected as a binary
tree structure,  which can be expressed as
\begin{eqnarray}
 &  & \mathcal{M}\left|q_{1},q_{2},\ldots,q_{N_{s}}\right\rangle \nonumber \\
 & = & \sum_{\left\{ l_{j,i}\right\} =1}^{D}\prod_{i=1}^{N_{s}}t_{1,i}\left(q_{i},q_{a_{i}},q_{b_{i}},q_{c_{i}},l_{2,i}\right)\nonumber \\
 &  & \times\left[\prod_{j=2}^{R}\prod_{i=1}^{N_{s}/2^{j-1}}t_{j,i}\left(l_{j,2i-1},l_{j,2i},l_{j+1,i}\right)\right]\nonumber \\
 &  & \times t_{R+1,1}\left(l_{R+1,1},l_{R+1,2}\right)\left|q_{1},q_{2},\ldots,q_{N_{s}}\right\rangle,
\end{eqnarray}
where $t_{1,i}$ is the leaf tensor (meaning the end
tensor of the tree structure) containing $4$ physical indices 
$q_{i},q_{a_{i}},q_{b_{i}},q_{c_{i}}$ 
of four sites on the plaquette with site $i$ on the top left vertex
and
$1$ virtual index, and $t_{2,i}$, $t_{3,i}$, $\ldots$, $t_{R+1,1}$ are
internal-node tensors which only contain virtual indices. 
Every $2$ neighboring leaf tensors share 2 common physical indices. 
In the TTN,
the number of leaf tensors connecting 2 neighboring sites is always
1, while it is 2 in the FTTN. 
The FTTN can reproduce the area law, because the TTN can be regarded as a subset of the FTTN with every leaf tensor contains only one physical index, and the TTN can reproduce the area law\cite{TTN2009}.
In the Monte Carlo sampling, the physical
degrees of freedom are fixed in every sample, so the tensor network
contraction in the FTTN will have the same computational complexity
as in the TTN. One can also increase the number of sites in every
leaf tensor to make further improvement.

It should be pointed out that regardless of the TTN or FTTN, the lattice
translational symmetry and rotational symmetry are broken. (Nevertheless,
that breaking is weaker than the MPS). To restore the symmetry, we
operate the quantum number projection by the summation over the spatially
translated and rotated tensor networks.

Besides tensor networks, one can also include backflow correlations\cite{Tocchio2008,Tocchio2011,Ido}
to further improve the correlation effect of the variational wave
function. The backflow correlations can be implemented in the pair-product
wave function\cite{Ido}, and in this case the additional computational cost of calculating
kinetic energy arises because the pairing amplitude with backflow correlations is dependent on a real space configuration of electrons,
which scales as $O\left(\gamma N_{s}^{2}\right)$. 
In the simplest consideration of the nearest neighbor sites backflow
correlations in two dimensions, the prefactor $\gamma$ is about $4000$. 
As a result,
the numerical cost of employment of backflow correlations is demanding
when calculation system size as large as $16\times16$. Therefore,
we do not implement backflow correlations in our calculation.

To further improve the accuracy of the VMC calculation, one can apply
the Lanczos method\cite{VMCLanczos1993}. After obtained optimized
wave function, we extend the wave function by multiplying the Hamiltonian
as 
\begin{equation}
\left|\psi_{n}\right\rangle =\left(1+\sum_{n=1}^{N}\alpha_{n}H^{n}\right)\left|\Psi\right\rangle ,
\end{equation}
where $\left(1+\sum_{n=1}^{N}\alpha_{n}H^{n}\right)$ can be regarded
as projection operator with variational parameters $\alpha_{n}$.
One can minimize the energy by choosing appropriate $\alpha_{n}$.
In principle, the accuracy can be systematically improved by increasing
$N$, but the computational cost grows exponentially. Therefore, we
employ Lanczos method up to the first step in our calculation.

\subsection{variational Monte Carlo}

We calculate the ground state by the variational Monte Carlo method
with the variational wave function Eq. (\ref{eq:TNVMC}) provided
in the last section. The expectation values of the Hamiltonian can
be calculated by Monte Carlo samplings of the real space electron
configurations $x$,

\begin{equation}
\left\langle H\right\rangle =\frac{\sum_{x}\left\langle \Psi\right|H\left|x\right\rangle \left\langle x\right|\left.\Psi\right\rangle }{\left\langle \Psi|\Psi\right\rangle }=\sum_{x}\rho\left(x\right)E\left(x\right)
\end{equation}
where

\begin{equation}
E\left(x\right)=\sum_{x'}\frac{\left\langle \Psi|x^{\prime}\right\rangle }{\left\langle \Psi|x\right\rangle }\left\langle x^{\prime}\left|H\right|x\right\rangle ,\rho\left(x\right)=\frac{\left\langle \Psi|x\right\rangle \left\langle x|\Psi\right\rangle }{\left\langle \Psi|\Psi\right\rangle },
\end{equation}
and $\rho$ is the weight in the importance sampling.

To find the ground state wave function, we optimize the variational
parameters by the SR method\citep{SR2001} in the variational Monte
Carlo. The SR method starts from an approximate power method
by the imaginary time evolution operator\cite{Ido}

\begin{equation}
e^{-\tau H}\approx1-\tau H.
\end{equation}

When $\tau$ is sufficiently small, it is reasonable to approximate
$\left(1-\tau H\right)\left|\overline{\Psi}\right\rangle $ as a linear combination
of the current wave function and its first derivatives,

\begin{equation}
\left(1-\tau H\right)\left|\overline{\Psi}\right\rangle \approx\left|\overline{\Psi}\right\rangle +\sum_{k=1}^{n_{p}}\gamma_{k}\left|\Psi^{k}\right\rangle ,
\end{equation}
where $\left|\overline{\Psi}\right\rangle=\left|\Psi\right\rangle/\sqrt{\left\langle \Psi|\Psi\right\rangle }$ 
and $\left|\Psi^{k}\right\rangle $ is the derivative of normalized
wave function with respect to a variational parameter $\alpha_{k}$

\begin{eqnarray}
\left|\Psi^{k}\right\rangle  & = & \frac{\partial}{\partial\alpha_{k}}\left(\frac{1}{\sqrt{\left\langle \Psi|\Psi\right\rangle }}\left|\Psi\right\rangle \right)\nonumber \\
 & = & \frac{1}{\sqrt{\left\langle \Psi|\Psi\right\rangle }}\left[\frac{\partial\left|\Psi\right\rangle }{\partial\alpha_{k}}-\frac{\left\langle \Psi\right|\partial/\partial\alpha_{k}\left|\Psi\right\rangle }{\left\langle \Psi|\Psi\right\rangle }\left|\Psi\right\rangle \right].
\end{eqnarray}
To determine the coefficients $\gamma_{k}$, we need to minimize the
cost function

\begin{eqnarray}
        f\left(\gamma_{k}\right) & = & \left\Vert \tau H\left|\overline{\Psi}\right\rangle +\sum_{k=1}^{n_{p}}\gamma_{k}\left|\Psi^{k}\right\rangle \right\Vert^{2} \nonumber \\
 & = & \tau^{2}\left\langle \overline{\Psi}\right|H^{2}\left|\overline{\Psi}\right\rangle +2\tau\sum_{k=1}^{n_{p}}\gamma_{k}\left\langle \overline{\Psi}\right|H\left|\Psi^{k}\right\rangle \nonumber \\
 &  & +\sum_{k=1}^{n_{p}}\sum_{l=1}^{n_{p}}\gamma_{l}\gamma_{k}\left\langle \Psi^{k}|\Psi^{l}\right\rangle ,
\end{eqnarray}
which can be obtained by computing the derivative with respect to
$\gamma_{k}$ and Let

\begin{equation}
\frac{\partial f\left(\gamma_{k}\right)}{\partial\gamma_{k}}=0.
\end{equation}
Then, $\gamma_{k}$ can be found by solving the following coupled
linear equation, 
\begin{equation}
\sum_{l=1}^{n_{p}}\left\langle \Psi^{k}|\Psi^{l}\right\rangle \gamma_{l}=-\tau\left\langle \overline{\Psi}\right|H\left|\Psi^{k}\right\rangle .\label{eq:SReq}
\end{equation}
Once we obtain the coefficients $\gamma_{k}$, the variational parameters
are updated as follows

\begin{equation}
\tilde{\alpha}_{k}=\alpha_{k}+\gamma_{k}.
\end{equation}
Then, we repeat these steps until the energy converges.

To avoid local minimum, we gradually decrease the step width $\tau$
and randomize the update of every parameter

\begin{equation}
\tau_{k}=-\Delta t\left(i\right)\eta\left(i,k\right),
\end{equation}
where $\Delta t\left(i\right)$ is a gradually reduced function of
SR step number $i$

\begin{equation}
\Delta t\left(i\right)=\xi^{i},
\end{equation}
and $\eta\left(i,k\right)$ is a random number in the interval of
$\left(r\left(i\right),1\right)$, with $r\left(i\right)=1-\left(1-r\left(0\right)\right)\nu^{i}$
to gradually reduce the randomness by selecting $\nu\in\left(0,1\right)$.

In order to optimize large number of parameters, one can solve the
SR equation iteratively by conjugate gradient (CG) method\cite{Neuscamman2012}.
The detailed implementation is described in Appendix \ref{Apdx:SRCG}.
The computational cost for solving the SR equation by CG scales as
$O\left(n_{s}n_{p}n_{\rm iter}\right)$ instead of $O\left(n_{s}n_{p}^{2}+n_{p}^{3}\right)$
needed for the solution of Eq.(\ref{eq:SReq}) by the simple matrix
inversion, where $n_{s}$ is the number of samples, $n_{p}$ is
the number of variational parameters and $n_{\rm iter}$ is the number of iterations in CG method. 
For comparison, the computational
cost for the energy calculation scales as 
$O\left(n_{s}\left(N_{s}^{3}+\log_{2}\left(N_{s}\right)N_{s}D^{3}\right)\right)$,
where $N_{s}^{3}$ comes from the calculation of Pfaffian, and $\log_{2}\left(N_{s}\right)N_{s}D^{3}$
comes from the contraction of tensor networks. Assume that we employ
the full sublattice, namely, $N_{{\rm sub}}=N_{s}$, in the variational
wave function, the number of parameters scales as $n_{p}\sim O\left(N_{s}^{2}+N_{s}D^{3}\right)$,
where $N_{s}^{2}$ comes from the Jastrow factor and $\phi_{{\rm pair}}$,
and $N_{s}D^{3}$ comes from the tensor network. Typically, we use $n_{\rm iter}\backsim O\left(10^{3}\right)$
in solving the SR equation by CG, so the dominant contribution
of computational cost comes from solving SR equation. In the calculation
of $16\times16$ square lattice, the computational cost of the VMC
part (with the CG) and the tensor network part are comparable when
$D=8$.

\section{\label{sec:benchmark}Benchmark results}

To benchmark our method, we test the Hubbard model defined in Eq.(\ref{Hubbard}).
\subsection{\label{Benchmark:1D} 1D Hubbard model}

\begin{figure}
\includegraphics[width=8cm]{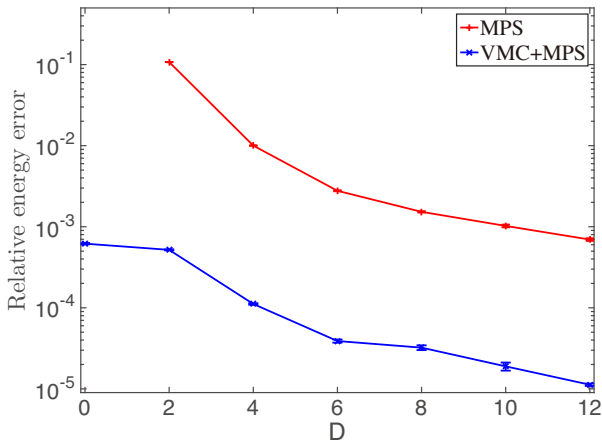} 
\caption{(color online) Relative errors of the ground state energy $\left|E-E_{{\rm exact}}\right|/\left|E\right|$
as a function of $D$ for the 1D Hubbard model with $L=16$, $U=10.0$,
and the number of electrons $N_{e}=10$. The red line is the conventional
MPS result, and the blue line is obtained by applying mVMC on the
variational wave function of Eq. (\ref{eq:TNVMC}) together with the
MPS for the tensor network part $\mathcal{M}$. In this calculation,
the full sublattice is employed with the spin projection. }
\label{fig:error_D_L16} 
\end{figure}

We first calculate the 1D Hubbard model with the periodic boundary
condition (PBC). Figure \ref{fig:error_D_L16} shows the $D$ dependence
of the relative error of the ground state energy with respect to the
exact result. The conventional MPS improves the accuracy about two
orders of magnitude from $D=2$ up to $D=12$. However, even with
$D=12$, the error is nearly $10^{-3}$. On the other hand, the conventional
mVMC method ($D=0$) already achieves the error slightly less than $10^{-3}$
and the combination with the MPS substantially improves the accuracy
with nearly the two orders of magnitude smaller error $\sim10^{-5}$
for $D=12$. By combining with the mVMC, the accuracy of MPS is enhanced
nearly two orders of magnitude.

\begin{figure}
\includegraphics[width=8cm]{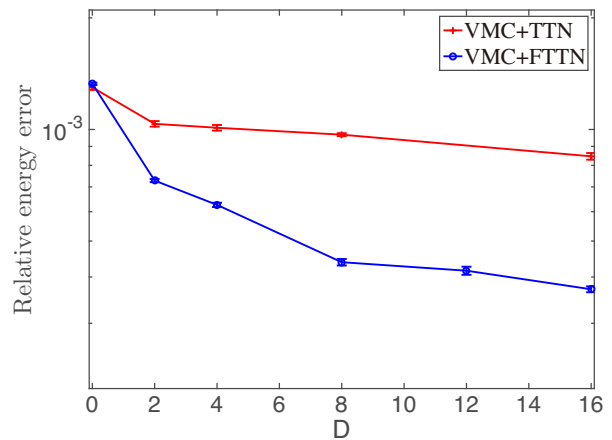} \caption{(color online) Relative errors of the ground state energy as a function
of $D$ in $4\times4$ 2D Hubbard model with $U=10.0$, and $N_{e}=10$.
The red and blue lines are obtained by the combination of the mVMC
with the TTN and FTTN, respectively. In this calculation, the full
sublattice is employed with the spin and space group projections.}
\label{fig:error_D_TTN_FTTN} 
\end{figure}

\subsection{\label{Benchmark:2D} 2D Hubbard model}

In the following, we calculate the 2D Hubbard model with the PBC.
It is necessary to rely on the Monte Carlo sampling to combine the VMC and the tensor network procedure. When the Monte Carlo sampling is introduced to the tensor network part, the initial (reference) wave function  $|\phi_{\rm ref}\rangle$ in the 2D case is required to be refined in advance as in the case of the VMC wavefunction. If we employ a simple state for the reference wavefunction such as $\sum_i|x_i \rangle$, which represents the equal-weight linear combination of all the real space basis function as employed in the conventional tensor network methods, the statistical error from the Monte Carlo sampling causes numerical instabilities.  In other words, the tensor network calculation is made possible only by combining with the VMC if the Monte Carlo sampling is employed to reduce the computational cost of the contraction in the tensor network.  Therefore, in the 2D case, we will not show the comparison with the solely FTTN result.
      
Figure \ref{fig:error_D_TTN_FTTN} shows the $D$ dependence of the
relative error of the ground state energy with respect to the exact
result. 
We find that both of the combination of the TTN and FTTN with mVMC improve
the accuracy of conventional mVMC results. In particular, the combination
with the FTTN shows more significant improvement than that with the
TTN. With the increase of the FTTN bond dimension $D$ up to 16, the
accuracy of the mVMC is improved by nearly $1$ order of magnitude.
Therefore, the combination of the FTTN and the mVMC provides a systematic
method to improve the accuracy of each method applied separately.

\begin{figure}
\includegraphics[width=8cm]{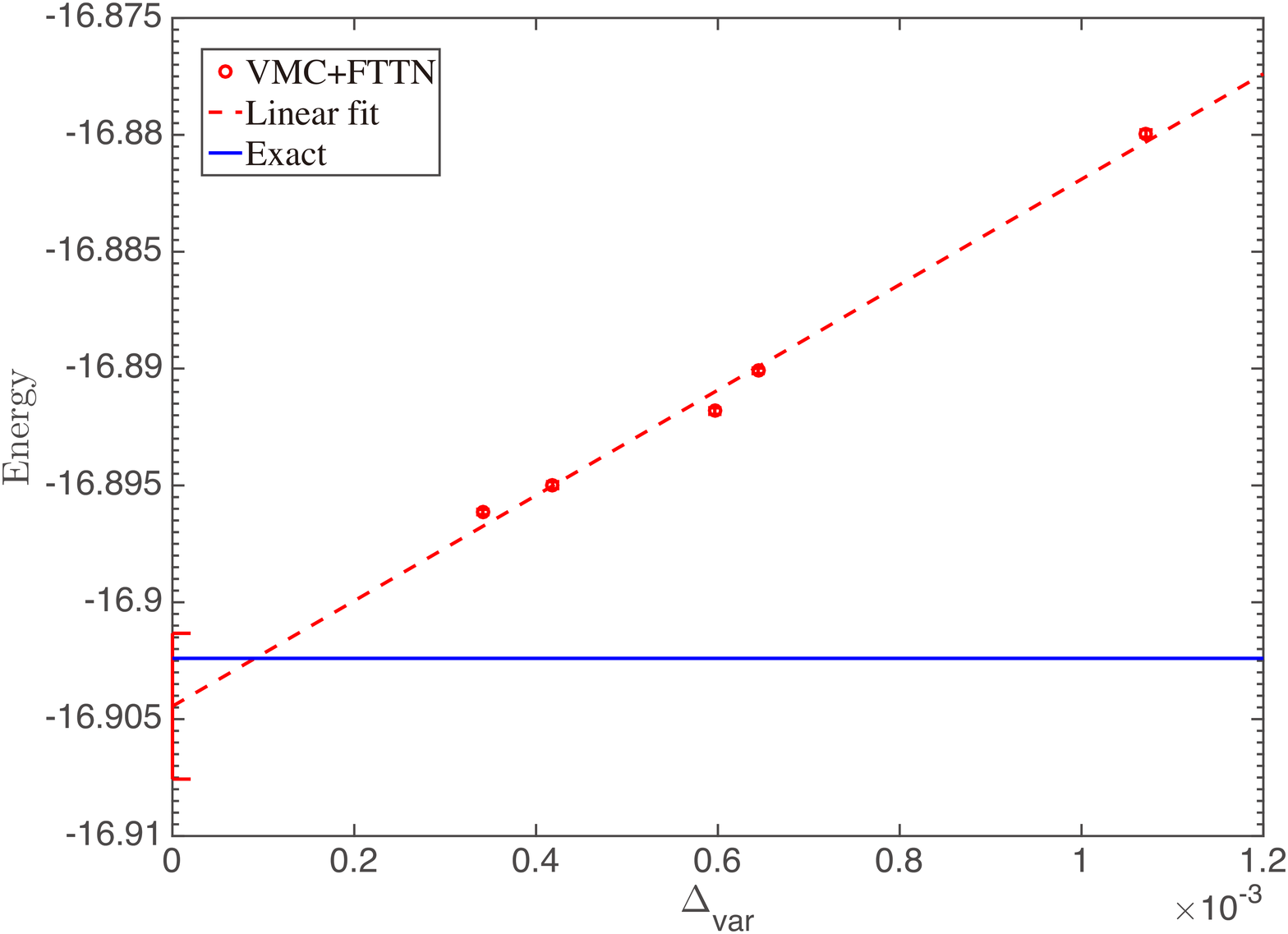} \caption{(color online) Variance dependence of energies for $D=0,2,4,8,16$
FTTN in $4\times4$ 2D Hubbard model with $U=10.0$, and $N_{e}=10$.
The red broken line represents the linear fitting of energies, and
the error bar on the $y$ axis is the fitting prediction bounds. The
blue line is the exact ground state energy.}

\label{fig:E_var} 
\end{figure}

In Fig. \ref{fig:E_var}, we plot the energy as a function of the
variance $\Delta_{{\rm var}}=\left(\left\langle H^{2}\right\rangle -\left\langle H\right\rangle ^{2}\right)/\left\langle H\right\rangle ^{2}$.
Since the energy is linearly proportional to $\Delta_{{\rm var}}$
for sufficiently small variance\cite{Kashima2000,Kashima2001,SR2001},
we can perform the linear fitting to extrapolate to the energy of
zero variance, so that more accurate ground state energy can be obtained.
Figure \ref{fig:E_var} shows that the extrapolated energy agrees
well with the exact result within the error bar of the linear fitting,
which indicates the order of the relative error as small as $\sim10^{-4}$. Though the accuracy is substantially improved, note that the strict variational principle satisfied before the extrapolation is not hold after the extrapolation, because of the possible extrapolation error.

\begin{figure}
\includegraphics[width=7.5cm]{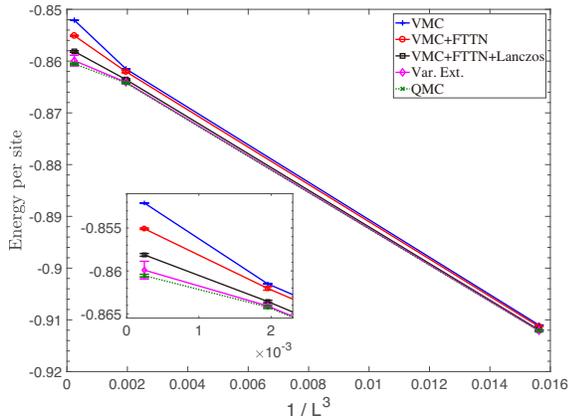} 
\caption{(color online) 
Lattice size dependence of the ground state energies
for 2D Hubbard model with $U=4.0$ at half filling. 
The blue line is the mVMC result without tensor network.
The red line represents the result of combination of the mVMC and FTTN with $D=16$. 
The black line is the result of the first step Lanczos. 
The magenta line is obtained from variance extrapolation.
The green dotted line represents the QMC result\cite{MPQin2016}.
The inset shows a magnified view of a portion of the
main figure. In this calculation, the full sublattice is employed
with the spin and space group projections.}

\label{fig:E_invL} 
\end{figure}

\begin{table}        
\caption{ Ground state energies per site
for 2D Hubbard model with $U=4.0$ at half filling 
for $8\times 8$ and $16\times 16$ square lattices.
The energies obtained from the first Lanczos step applied to the combined
mVMC and FTTN, the variance extrapolation, and the QMC\cite{MPQin2016}, 
which are plotted in Fig.~\ref{fig:E_invL}, are listed. 
The parentheses denote the error bars in the last digit.}
\begin{ruledtabular}
\begin{tabular}{cccc}
& VMC+FTTN+Lanczos & Var. Ext. & QMC\cite{MPQin2016} \\ 
\hline
$8 \times 8$ & $-0.8636(2)$& $-0.8641(2)$ & $-0.8642(2)$  \\
$16 \times 16$ & $-0.8581(2)$& $-0.860(1)$ & $-0.8605(2)$ \\
\end{tabular}
\end{ruledtabular}
\label{table:E_invL}
\end{table}

We now perform the calculation for larger system sizes. Figure \ref{fig:E_invL}
shows the ground state energies for 2D Hubbard model with $U=4.0$
at half filling on $4\times4$, $8\times8$ and $16\times16$ square
lattices with the periodic-antiperiodic boundary conditions. 
We show comparisons among the mVMC, mVMC combined with FTTN, 
the first Lanczos step applied to the combined mVMC and FTTN, the variance extrapolation, and the QMC results.
Our calculated energies agree well with the QMC results, 
which is expected to be practically exact. 
The relative error with respect to the QMC energies\cite{MPQin2016}
is about $0.3\%$ on the $16\times16$ lattice size after the first-step
Lanczos operation, and the extrapolated energies agree well with the QMC result within the error bar of the linear fitting. 
The first Lanczos step applied to the combined
mVMC and FTTN, the variance extrapolation, and the QMC
energies for the $8 \times 8$ and $16 \times 16$ square lattices are listed in Table~\ref{table:E_invL}.

\section{doped Hubbard model on square lattice\label{dopedHubbard}}

In this section we show applications of the present method to the carrier doped
Hubbard model on the square lattice to gain insight into a long-standing issue of
the high-$T_{\rm c}$ superconductivity and severe competitions among various orders
and fluctuations, given that the Hubbard model captures some essential physics of the high-$T_{\rm c}$ superconductivity.
We show comparisons among the mVMC, mVMC combined with FTTN and  the first Lanczos step applied to the combined mVMC and FTTN.   

\subsection{Energy\label{dopedHubbardEnergy}}

\begin{figure*}
\includegraphics[width=16cm]{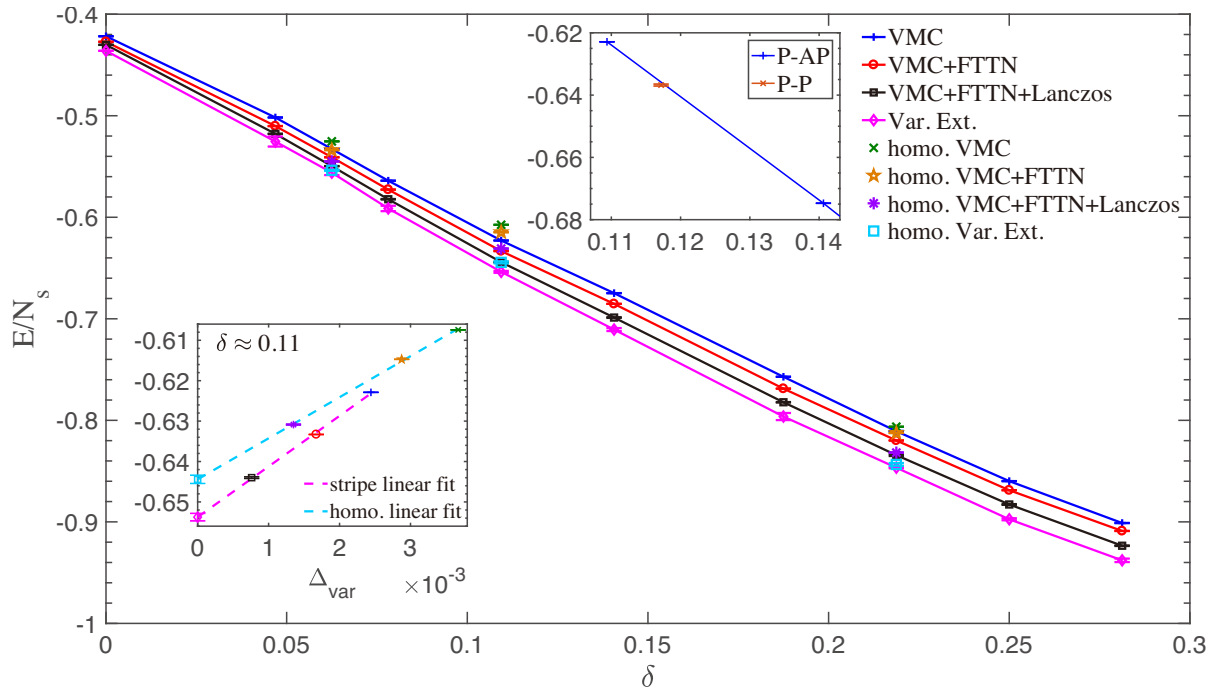} 
\caption{(color online) Hole doping concentration ($\delta$) dependence of the ground state energy
per site for $16\times16$ square lattice of 2D square lattice Hubbard model at $U=10$ with the periodic-antiperiodic boundary condition. 
The blue line is the mVMC
result without tensor network. The red line represents the result
of combination of the mVMC and FTTN with $D=16$. The black line is
the result of the first step Lanczos, and the magenta line is obtained
from variance extrapolation. 
In this calculation, we have performed the optimization from
the initial wave function with the optimized period of the stripe
order coexisting with the $d$-wave superconductivity and employ the
$16\times2$ sublattice for $f_{ij}$, which allows various charge/spin
orders.
For the doping smaller than $15\%$, the ground state spin
stripe period is $16$, and charge stripe period is $8$, while for the
doping larger than $15\%$, the spin stripe period is $8$, and the
charge stripe period is $4$, irrespective of the methods. 
In addition to the ground states, we show metastable excited states obtained from the optimization performed
from homogeneous superconducting initial wave function: 
the green crosses are the mVMC results without tensor network, 
the orange pentagrams represent the results of the combination
of the mVMC and FTTN with $D=16$, 
both of which preserves the charge homogeneity even after the optimization, the purple stars are the results of the first step Lanczos, 
and the cyan squares are obtained from variance extrapolation estimated in the way shown in the lower inset (see below).
The upper inset shows
the energy per site of the mVMC results with periodic-periodic boundary
condition (red point) and periodic-antiperiodic boundary condition
(blue line).
The lower inset shows the variance dependence of energies
for the mVMC, combination of the mVMC and FTTN with $D=16$, 
and the first step Lanczos with stripe and homogeneous initial states at $\delta \sim 0.11$,
in which the broken lines represent the linear fitting of energies,
and the energy extrapolation to zero variance is represented as the magenta diamond and cyan square for stripe and homogeneous initial states respectively.
}
\label{fig:E_doping} 
\end{figure*}

\begin{table*}        
\centering

\caption{ Ground state energies per site of 
$16\times 16$ 2D square lattice Hubbard model at $U=10$ 
with the periodic-antiperiodic boundary condition for various doping concentrations $(\delta)$. 
The energies obtained from the first Lanczos step applied to the combined mVMC and FTTN, and the variance extrapolation
with stripe and homogeneous initial states 
which are plotted in Fig. ~\ref{fig:E_doping}, are listed.
The parentheses denote the error bars in the last digit.}
\begin{tabular}{ccccc}
\hline
\hline
$\delta$ & VMC+FTTN+Lanczos & Var. Ext.  &
homo. VMC+FTTN+Lanczos & homo. Var. Ext. \\
\hline
$0.00$ & $-0.43018(6)$& $-0.4361(1)$ &  &  \\
$0.05$ & $-0.51782(8)$& $-0.525(5)$ & & \\
$0.06$ & $-0.54963(7)$& $-0.556(2)$ & $-0.54497(6)$ & $-0.5537(6)$\\
$0.08$ & $-0.5822(2)$& $-0.591(3)$ & &\\
$0.11$ & $-0.6442(4)$& $-0.6538(9)$ &  $-0.6309(1)$ & $-0.644(1)$ \\
$0.14$ & $-0.6988(3)$& $-0.711(1)$ & &\\
$0.19$ & $-0.7822(2)$& $-0.796(3)$ &  & \\
$0.22$ & $-0.8343(9)$& $-0.8466(6)$ & $-0.8318(2)$ & $-0.843(1)$\\
$0.25$ & $-0.8829(4)$& $-0.897(1)$ &  & \\
$0.28$ & $-0.9235(2)$& $-0.938(2)$ &  & \\
\hline
\hline
\end{tabular}
\label{table:E_doping}
\end{table*}

\begin{figure}
\includegraphics[width=8cm]{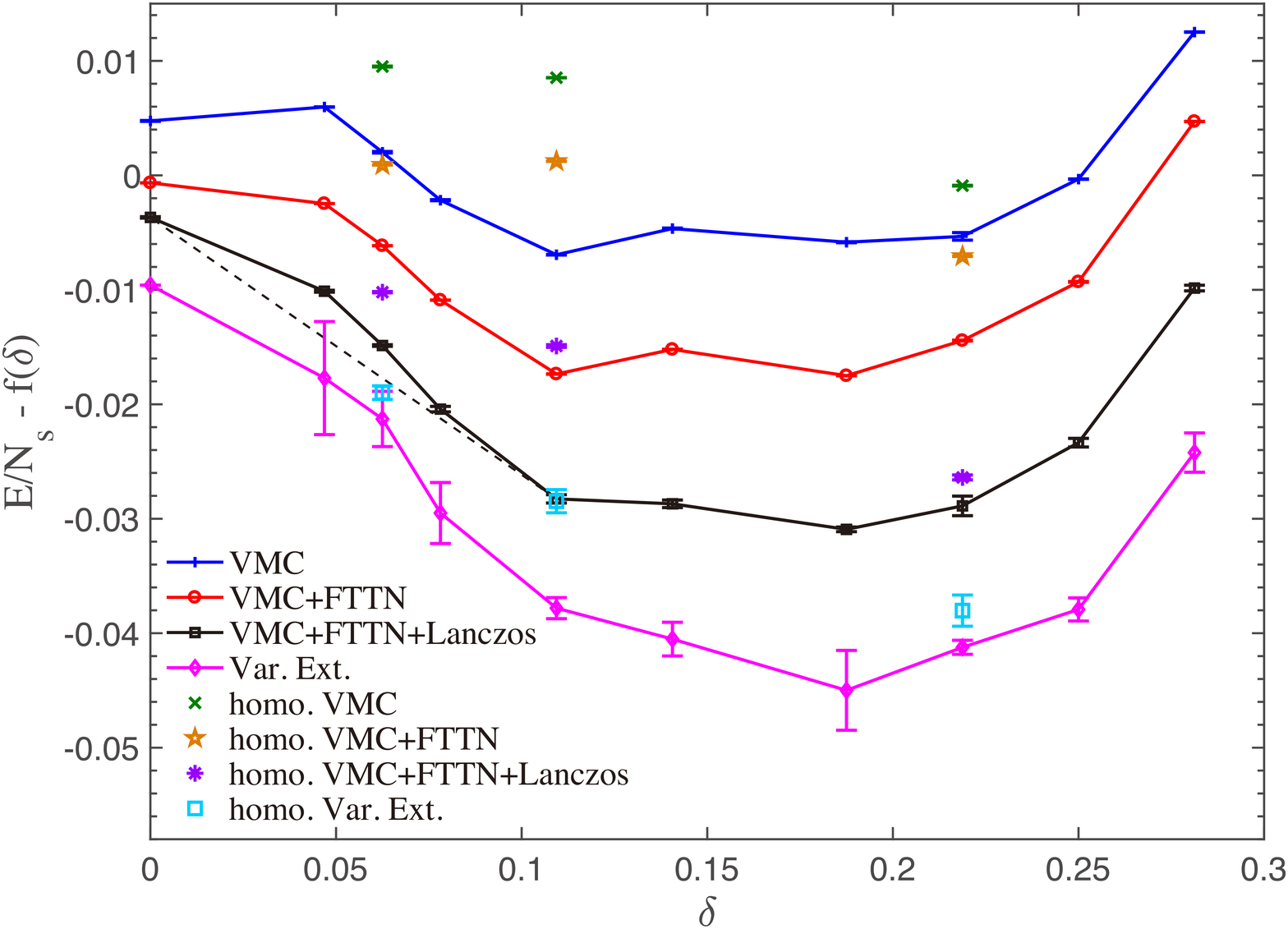} \caption{(color online) Hole doping concentration ($\delta$) dependence of
the ground state energy per site for the Hubbard model on the $16\times16$
square lattice at $U=10$.  
The data are the same as Fig.~\ref{fig:E_doping}, but here, a linear function $f(\delta)=-1.7297\delta-0.4270$
has been subtracted from the energy to enhance the visibility. Notations are the same as Fig.~\ref{fig:E_doping}.
The black dashed line indicates the range of the phase separation
near half filling ($0<\delta \lesssim 0.11$) for the example of VMC+FTTN+Lanczos.}
\label{fig:E_doping_ft} 
\end{figure}

Figure \ref{fig:E_doping} shows the
doping concentration dependence of the ground state energy on $16\times16$ square
lattice for $U=10.0$. The energy difference between periodic-periodic boundary
and periodic-antiperiodic boundary results are negligible on $16\times16$
lattice size (See upper inset of Fig. \ref{fig:E_doping}), so we have performed
all the calculation with the periodic-antiperiodic boundary condition
in this section. 
We employ the $16 \times 2$ sublattice in the variational wave function
to be compatible with the possible stripe orders.
The result shows that the combination of the FTTN substantially lower the energy of the corresponding mVMC result. 
We have performed the optimization from a homogeneous $d$-wave
superconducting state and a stripe order coexisting with $d$-wave
superconducting order, and we find that the stripe ordered state coexisting
with the weak $d$-wave superconductivity provides lower energy, while 
the state optimized from the homogeneous $d$-wave
superconducting state stays metastable as an excited state at least for $\delta<0.25$.
The energies of the first Lanczos step applied to the combined
mVMC and FTTN, and the variance extrapolation,
which are obtained from stripe and homogeneous initial states, 
are listed in Table~\ref{table:E_doping}. 

In Fig. \ref{fig:E_doping_ft}, we show the same data as the main panel of Fig. \ref{fig:E_doping}, but in an enlarged scale of the vertical axis after subtracting a linear function $f(\delta)$ to emphasize the difference among different methods. Note that the phase separation region
determined by drawing the tangent line from the point at $\delta=0$
as shown in the dashed line for the VMC+FTTN+Lanczos data suggests that the phase separation for $0\leq\delta\lesssim0.1$,
which is narrower than the phase separation region $0\leq\delta\lesssim0.2$
in Ref.~\onlinecite{Misawa2014PRB}, while below $\delta\sim 0.1$, the survival of the phase separation is robust. The difference of the present result from Ref.~\onlinecite{Misawa2014PRB} is ascribed to
the fact that the present calculation allows the finite-period phase
separation, namely the stripe order, replacing the phase separation,
which is regarded as the ``infinite period\char`\"{} charge order.
The charge/spin stripe order suppresses the $d$-wave superconducting
correlation as we see below.

\subsection{$d$-wave Superconducting Correlation \label{dopedHubbardSC}}
\begin{figure}
\includegraphics[width=8cm]{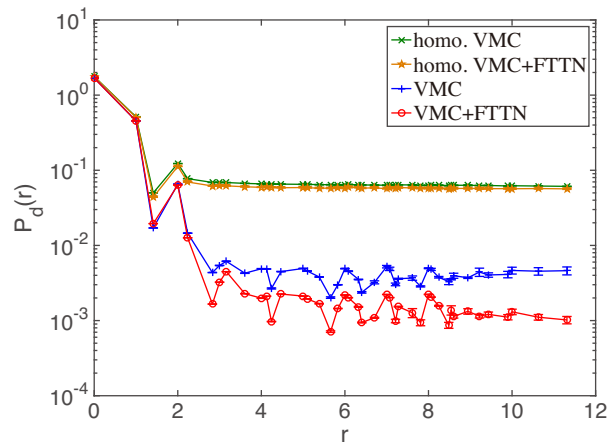} 
\caption{(color online) Distance ($r$) dependence of superconducting correlation
for $\delta\sim0.11$ on $16\times16$ lattice with $U=10$. The blue line is the mVMC result,
and the red line is the result of the combination of the mVMC and $D=16$
FTTN for the ground states, which are obtained by starting from the stripe order coexisting with the $d$-wave
superconductivity initial wave function. We also plot the correlation for the excited states, where the green
line is the mVMC result, and the orange line is the result of combination
of the mVMC and $D=16$ FTTN, obtained by starting from the homogeneous $d$-wave
superconducting initial wave function, which results in the metastable excited states after the optimization as we see in Figs.~\ref{fig:E_doping} and \ref{fig:E_doping_ft}. Here, for a given distance, the maximum value of the correlation is plotted. }

\label{fig:sc_r} 
\end{figure}

\begin{figure}
\includegraphics[width=8cm]{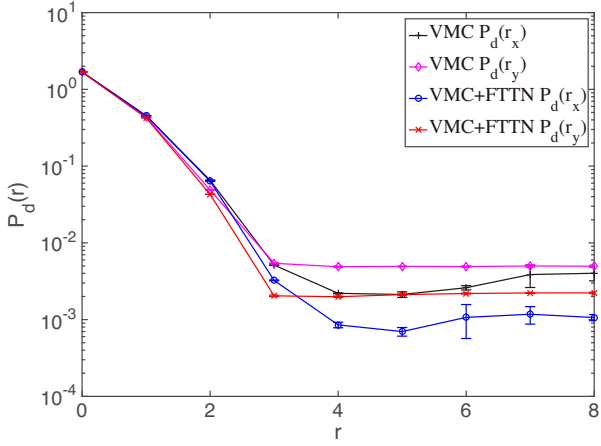} \caption{(color online) Distance ($r$) dependence of superconducting correlation
along $x$ (blue circles and black +signs) and $y$ (red crosses and magenta diamonds)
directions for $\delta\sim0.11$ on $16\times16$ lattice with $U=10$.
The black and magenta lines are the mVMC result without tensor network.
The red and blue lines represent the result of the combination of the
mVMC and the $D=16$ FTTN. The stripe direction is along the $y$ direction.}

\label{fig:sc_rxy} 
\end{figure}

In Fig. \ref{fig:sc_r}, we show the superconducting correlations
defined as
\begin{eqnarray}
P_{d}\left(\mathbf{r}\right) & = & \frac{1}{2N_{s}}\sum_{\mathbf{r}_{i}}\left\langle \Delta_{d}^{\dagger}\left(\mathbf{r}_{i}\right)\Delta_{d}\left(\mathbf{r}_{i}+\mathbf{r}\right)+\right.\nonumber \\
 &  & \left.\Delta_{d}\left(\mathbf{r}_{i}\right)\Delta_{d}^{\dagger}\left(\mathbf{r}_{i}+\mathbf{r}\right)\right\rangle, 
 \label{Pd}
\end{eqnarray}
where
\begin{equation}
\Delta_{d}\left(\mathbf{r}_{i}\right)=\frac{1}{\sqrt{2}}\sum_{\mathbf{r}}f_{d_{x^{2}-y^{2}}}\left(c_{\mathbf{r}_{i}\uparrow}c_{\mathbf{r}_{i}+\mathbf{r}\downarrow}-c_{\mathbf{r}_{i}\downarrow}c_{\mathbf{r}_{i}+\mathbf{r}\uparrow}\right),
\end{equation}
and $f_{d_{x^{2}-y^{2}}}$ is the $d_{x^{2}-y^{2}}$ superconducting
pairing symmetry factor
\begin{equation}
f_{d_{x^{2}-y^{2}}}\left(\mathbf{r}\right)=\delta_{r_{y},0}\left(\delta_{r_{x},1}+\delta_{r_{x},-1}\right)-\delta_{r_{x},0}\left(\delta_{r_{y},1}+\delta_{r_{y},-1}\right),
\end{equation}
where $\mathbf{r}=\left(r_{x},r_{y}\right)$. Both of the results for the ground states with the coexisting stripe and superconductivity as well as the excited states with charge uniform superconducting states presented in Figs.~\ref{fig:E_doping} and \ref{fig:E_doping_ft} are shown.  The both results show that
the combination of the FTTN slightly suppresses the superconducting
correlation on the long range part in comparison to the mVMC results.

Since the ground state we obtained has a stripe order, we show the
superconducting correlation along $x$ and $y$ directions separately for the ground state
in Fig. \ref{fig:sc_rxy}. We see that the combination of the
FTTN suppresses the superconducting correlation both along $x$ and
$y$ directions with factors two to three, and the superconducting correlation along the $y$
direction, which is the stripy direction, shows larger long range
correlation than along the $x$ direction, indicating that the charge
modulation suppresses the superconductivity as expected.
In contrast the long-ranged part of the amplitude of the superconducting order is more than one order of magnitude larger for the charge homogeneous excited states. It should be noted that, even for the stripe direction in the hole rich region, the superconducting correlation is much lower than the case of the charge homogeneous states.  Since the long-ranged part of $P_d$ is the square of the order parameter, the order parameter $\langle \Delta_d\rangle$ is more than factor 3 larger for the charge-uniform excited states. It suggests that the superconductivity can be substantially enhanced from the equilibrium ground state if one can keep the metastable charge-uniform state.  Nevertheless, despite weak order, the charge-inhomogeneous ground state preserves the saturated long-ranged correlation particularly in the stripe direction ($y$ direction).  
In the direction crossing the stripe ($x$ direction), the correlation shows the long-ranged saturation to a smaller value with oscillation with the period of the charge stripe.  
Since the charge-stripe long-range order may be sensitively destroyed by the randomness such as that caused by the distribution of the dopant atoms in the real compounds of the cuprate superconductors and may be replaced with domain structures,
the long-range superconducting order may further be suppressed than the values in the present ideally periodic stripe order.  

\begin{figure}
\includegraphics[width=8cm]{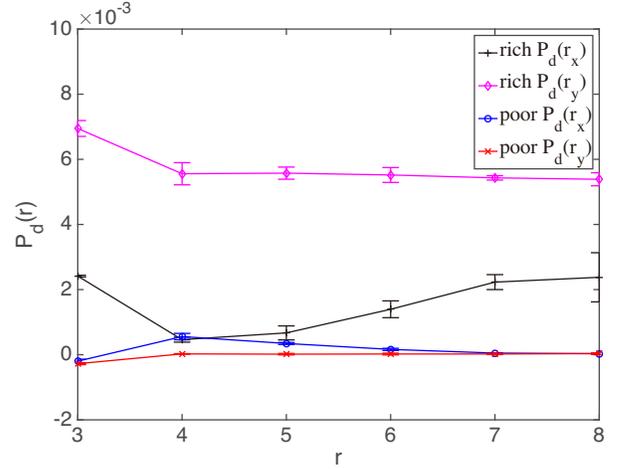} 
\caption{(color online) Distance ($r$) dependence of superconducting correlation
along $x$ (blue circles and black +signs) and $y$ (red crosses and magenta diamonds)
directions for $\delta\sim0.11$ on $16\times16$ lattice with $U=10$
plotted by taking the origin of the correlation at the minimum (blue circles and red crosses) and maximum (black +signs and magenta diamonds) columns of the hole density.
Here only the results calculated by the combination of mVMC and FTTN with $D=16$ are shown.
 The stripe direction is along the $y$ direction.}
\label{fig:sc_xy_Ne228_holerich} 
\end{figure}

Here in Fig.~\ref{fig:sc_xy_Ne228_holerich}, we show the superconducting correlation by taking the origin $\mathbf{r}_{i}$ in Eq. (\ref{Pd}) at the maximum and minimum values of the hole density (namely, columns of the stripe with smallest (largest) electron densities).  It clearly shows that the superconducting correlation along the stripe direction stays large at long distance only for the hole-rich columns while that at the hole poor region is extremely small. This is because the hole-poor region is essentially Mott insulating like (see the data for the charge density discussed below).  The correlation in the $x$ direction shows an oscillatory behavior and it confirms that the hole-rich superconducting strips are bridged each other essentially by the mechanism of the Josephson junction through the hole-poor strips, which is the reason why the superconducting order stays smaller than the charge homogeneous state. 

\subsection{Spin and Charge Correlations \label{dopedHubbardSpinCharge}}

To identify the stripe order in the ground state, we show the spin structure factor
\begin{equation}
S_{s}\left(\mathbf{k}\right)=\frac{1}{3N_{s}}\sum_{\mathbf{r},\mathbf{r^{\prime}}}
\left\langle s_{\mathbf{r}}^{z}s_{\mathbf{r+r^{\prime}}}^{z}\right\rangle e^{i\mathbf{k}\cdot\mathbf{r^{\prime}}},
\end{equation}
where $s_{\mathbf{r}}^{z}=n_{\mathbf{r}\uparrow}-n_{\mathbf{r}\downarrow}$.
We also present the charge structure factor
\begin{equation}
S_{c}\left(\mathbf{k}\right)=\frac{1}{N_{s}}\sum_{\mathbf{r},\mathbf{r^{\prime}}}
\left\langle n_{\mathbf{r}}n_{\mathbf{r+r^{\prime}}}\right\rangle e^{i\mathbf{k}\cdot\mathbf{r^{\prime}}},
\end{equation}
where $n_{\mathbf{r}}=n_{\mathbf{r}\uparrow}+n_{\mathbf{r}\downarrow}$.

\begin{figure}
\includegraphics[width=8cm]{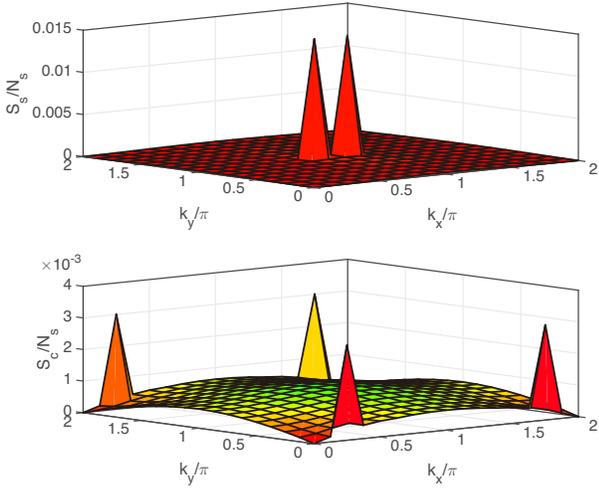} \caption{(color online) Spin (upper panel) and charge (lower panel) structure
factors at $\delta\sim0.11$ on $16\times16$ lattice with $U=10$.
}
\label{fig:SF_Ne228} 
\end{figure}

Figure \ref{fig:SF_Ne228} shows the spin and charge structure factor
at $\delta\sim0.11$. The peak of spin structure factor is at $\left(\frac{7\pi}{8},\pi\right)$,
and the peak of charge structure factor is at $\left(\frac{\pi}{4},0\right)$,
which indicates that the ground state has a stripe order 
with $(l_c, l_s)=(8,16)$, where $l_c$ ($l_s$) denotes the charge (spin) wave length in a stripe phase.

\begin{figure}
\includegraphics[width=8cm]{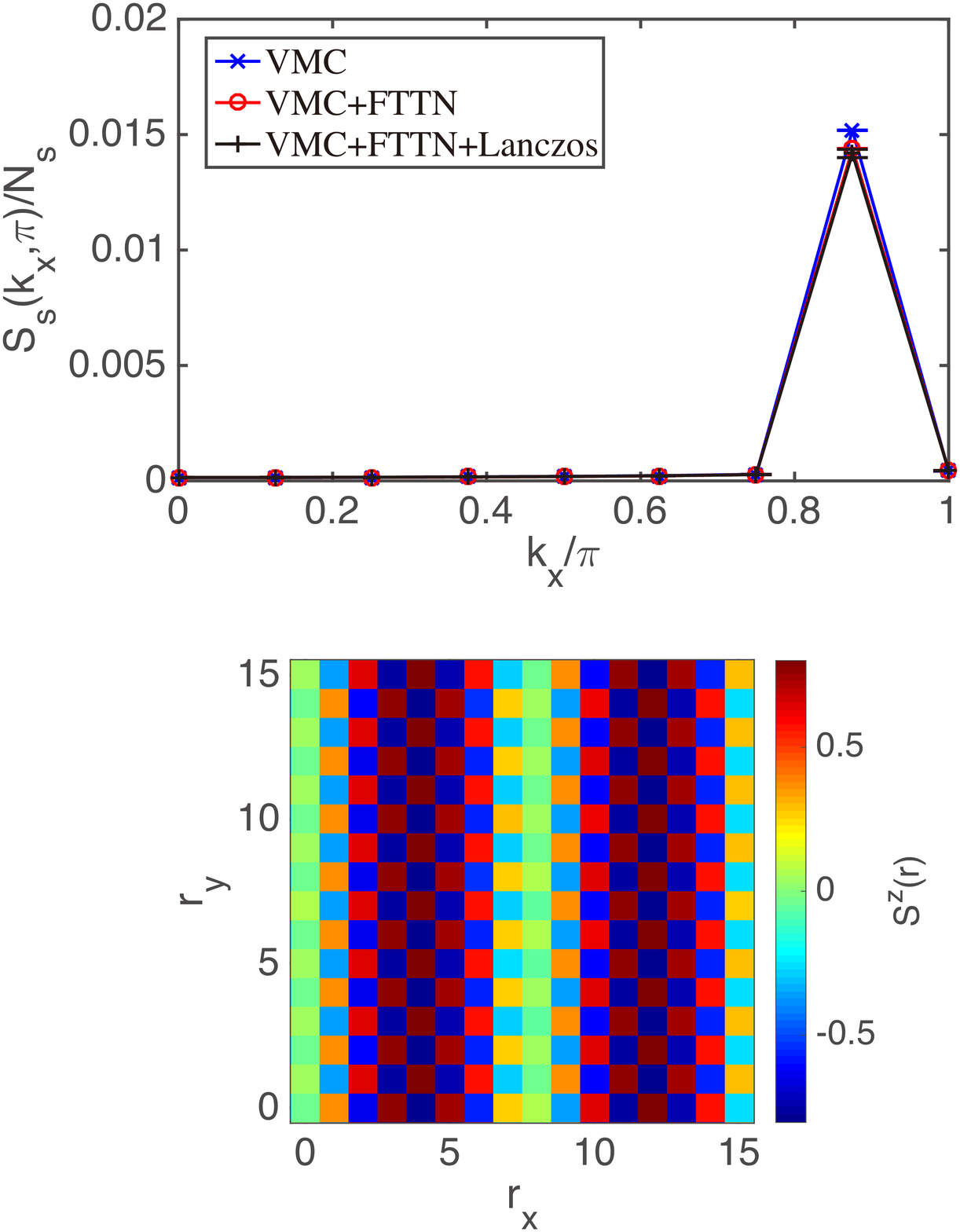} \caption{(color online) Spin structure factor at $k_y=0$ (upper panel) and spin configuration
(lower panel) at $\delta\sim0.11$ on $16\times16$ lattice with $U=10$.
In the upper panel, the blue line is the mVMC result, the red line
is the result by combining the mVMC and the FTTN with $D=16$, and
the black line is with the first Lanczos step. $S_{s}\left(\mathbf{k}\right)$
has a peak at $\mathbf{k=}\left(\frac{7}{8}\pi,\pi\right)$.}

\label{fig:SDW_Ne228} 
\end{figure}

\begin{figure}
\includegraphics[width=8cm]{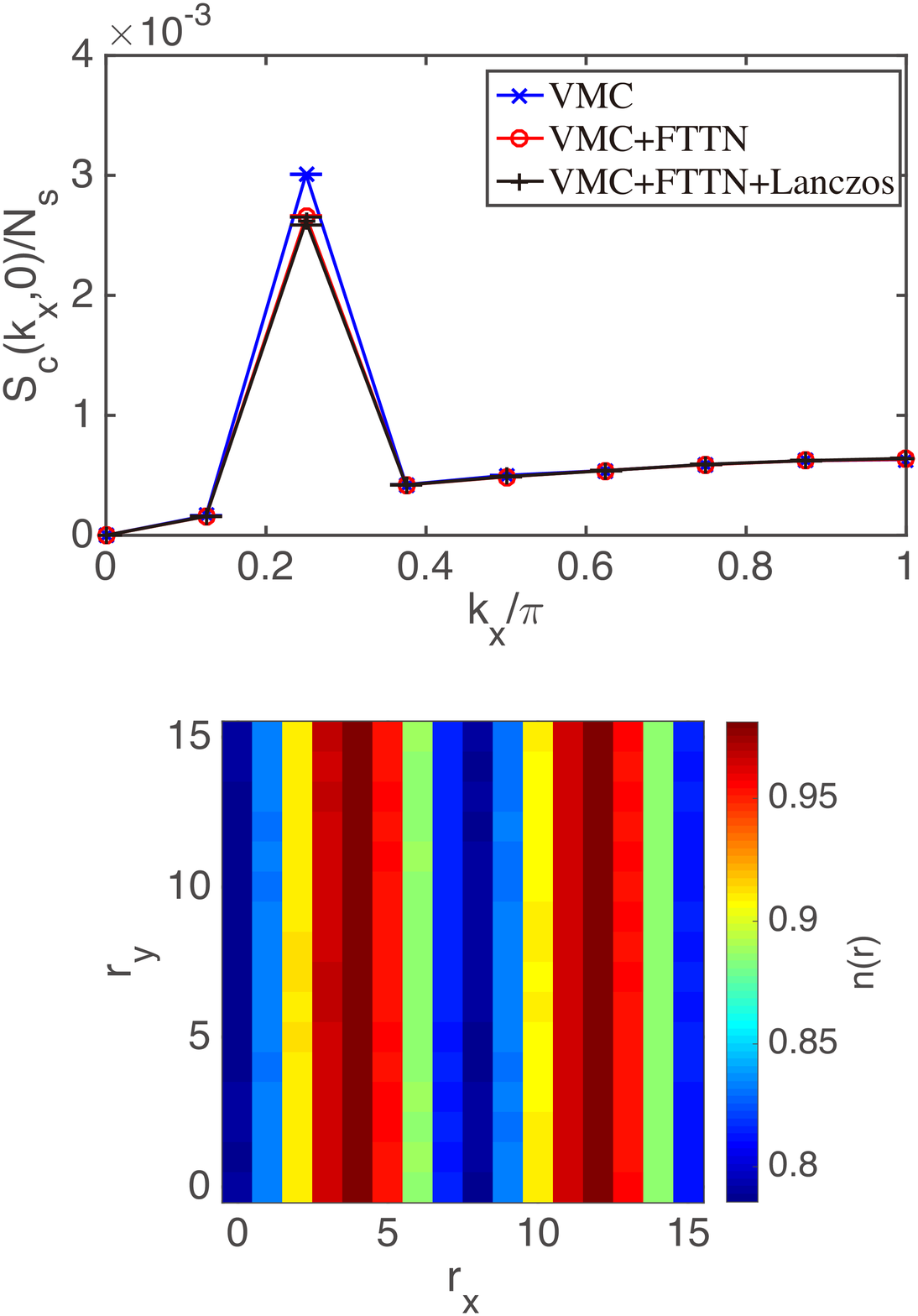} 
\caption{(color online) Charge structure factor (upper panel) and charge configuration
(lower panel) at $\delta\sim0.11$ on $16\times16$ lattice with $U=10$.
In the upper panel, the blue line is the mVMC result, the red line
is the result of combining the mVMC and the FTTN with $D=16$, and
the black line is with the first Lanczos step. $S_{c}\left(\mathbf{k}\right)$
has a peak at $\mathbf{k=}\left(\frac{1}{4}\pi,0\right)$.}

\label{fig:CDW_Ne228} 
\end{figure}

Then, we compare the stripe order obtained by different numerical
methods in Fig. \ref{fig:SDW_Ne228} and Fig. \ref{fig:CDW_Ne228}.
Our calculation shows that the combined mVMC, FTTN and
first Lanczos step very slightly lower the peak of the spin structure factor in comparison to the VMC+FTTN results.
The difference between the VMC and VMC+FTTN results is also small.

Although the true ground state of a finite-size system must preserves the translational symmetry, in our calculated results of the stripe ordered states,  
the translational symmetry is explicitly broken if the momentum projection is not imposed, because the system size is fairly large.  To show the spin and charge stripe patterns in the real space, we have computed the local spin
density along $z$ direction
\begin{equation}
S^{z}\left(\mathbf{r}\right)=\left\langle n_{\mathbf{r}\uparrow}-n_{\mathbf{r}\downarrow}\right\rangle ,
\end{equation}
and the local charge density 
\begin{equation}
n\left(\mathbf{r}\right)=\left\langle n_{\mathbf{r}\uparrow}+n_{\mathbf{r}\downarrow}\right\rangle,
\end{equation}
which are shown in the color scale plot of the spin and charge configuration
in the lower panel of Fig. \ref{fig:SDW_Ne228} and Fig. \ref{fig:CDW_Ne228},
respectively.
We note that the stripe order has the amplitude as large as 0.2, 
implying that the charge modulation extends from the Mott insulating density ($\delta \sim 0$) to $\delta\sim 0.2$.
In the realistic condition with the potential randomness and long-range Coulomb interaction, 
this amplitude of the long range order may be weakened.

\begin{figure}
\includegraphics[width=8cm]{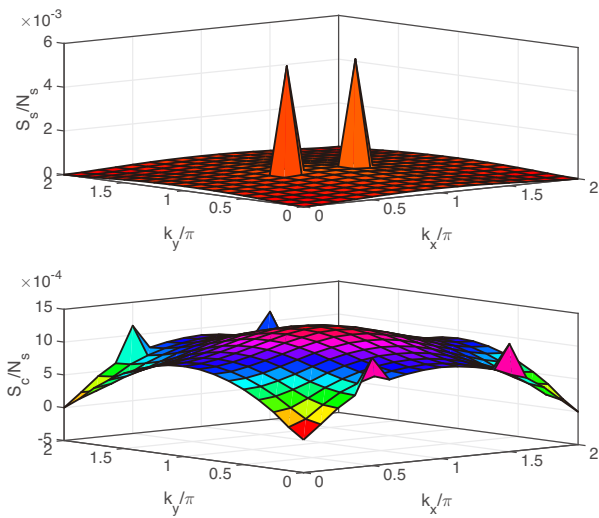} \caption{(color online) Spin (upper panel) and charge (lower panel) structure
factors at $\delta\sim0.22$ on $16\times16$ lattice with $U=10$.}

\label{fig:SF_Ne200} 
\end{figure}

In the overdoped region, the ground state may show a different stripe
order. We calculate the spin and charge structure factor at $\delta\sim0.22$,
which are shown in Fig. \ref{fig:SF_Ne200}. The peak of spin structure
factor is at $\left(\frac{3\pi}{4},\pi\right)$, and the peak of the
charge structure factor is at $\left(\frac{\pi}{2},0\right)$, which
indicate that the ground state has a stripe order with $(l_c, l_s)=(4,8)$. The reduction of the period with the increasing hole concentration is intuitively understood from the deceasing mean hole distance with doping and also consistent with the experimental indications of the diffuse peak in neutron scattering in the cuprates~\cite{Yamada1998,Dai1998}.

\begin{figure}
\includegraphics[width=8cm]{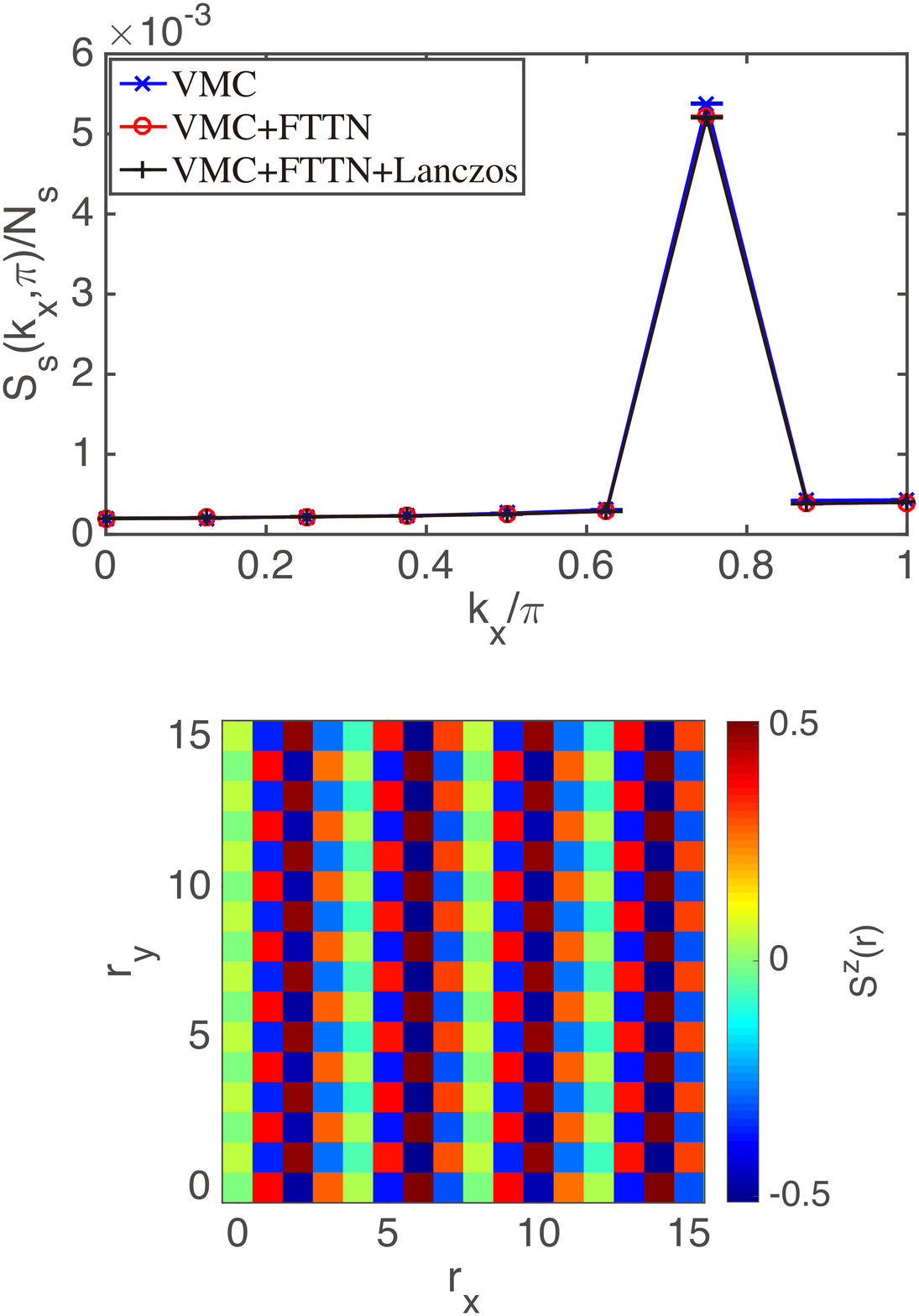} \caption{(color online) Spin structure factor (upper panel) spin configuration
(lower panel) at $\delta\sim0.22$ on $16\times16$ lattice with $U=10$.
In the upper panel, the blue line is the mVMC result, the red line
is the result of combining the mVMC and the FTTN with $D=16$, and
the black line is with the first Lanczos step. $S_{s}\left(\mathbf{k}\right)$
has a peak at $\mathbf{k=}\left(\frac{3}{4}\pi,\pi\right)$.}

\label{fig:SDW_Ne200} 
\end{figure}

\begin{figure}
\includegraphics[width=8cm]{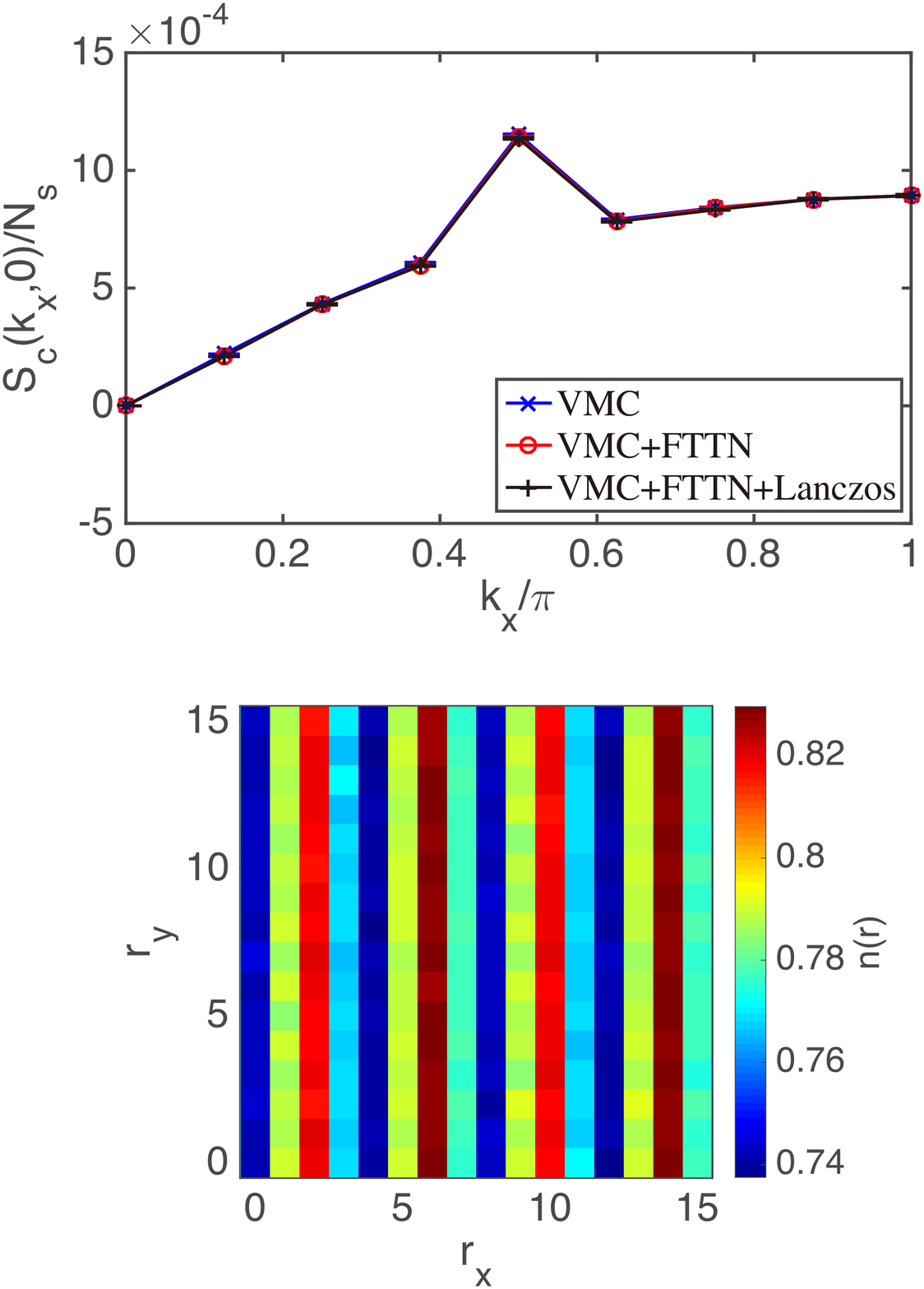} \caption{(color online) Charge structure factor (upper panel) and charge configuration
(lower panel) at $\delta\sim0.22$ on $16\times16$ lattice with $U=10$.
In the upper panel, the blue line is the mVMC result, the red line
is the result of combining the mVMC and the FTTN with $D=16$, and
the black line is with the first Lanczos step. $S_{c}\left(\mathbf{k}\right)$
has a peak at $\mathbf{k=}\left(\frac{1}{2}\pi,0\right)$.}

\label{fig:CDW_Ne200} 
\end{figure}

From Figs. \ref{fig:SDW_Ne200} and \ref{fig:CDW_Ne200}, we can
see that the combination of the mVMC with the FTTN and first Lanczos
step provide nearly the same structure factor. The spin and charge
stripe patterns can be seen in the color scale plots of the spin and
charge configurations in the lower panels of Fig. \ref{fig:SDW_Ne200}
and Fig. \ref{fig:CDW_Ne200}, respectively.

\begin{figure*}
\includegraphics[width=16cm]{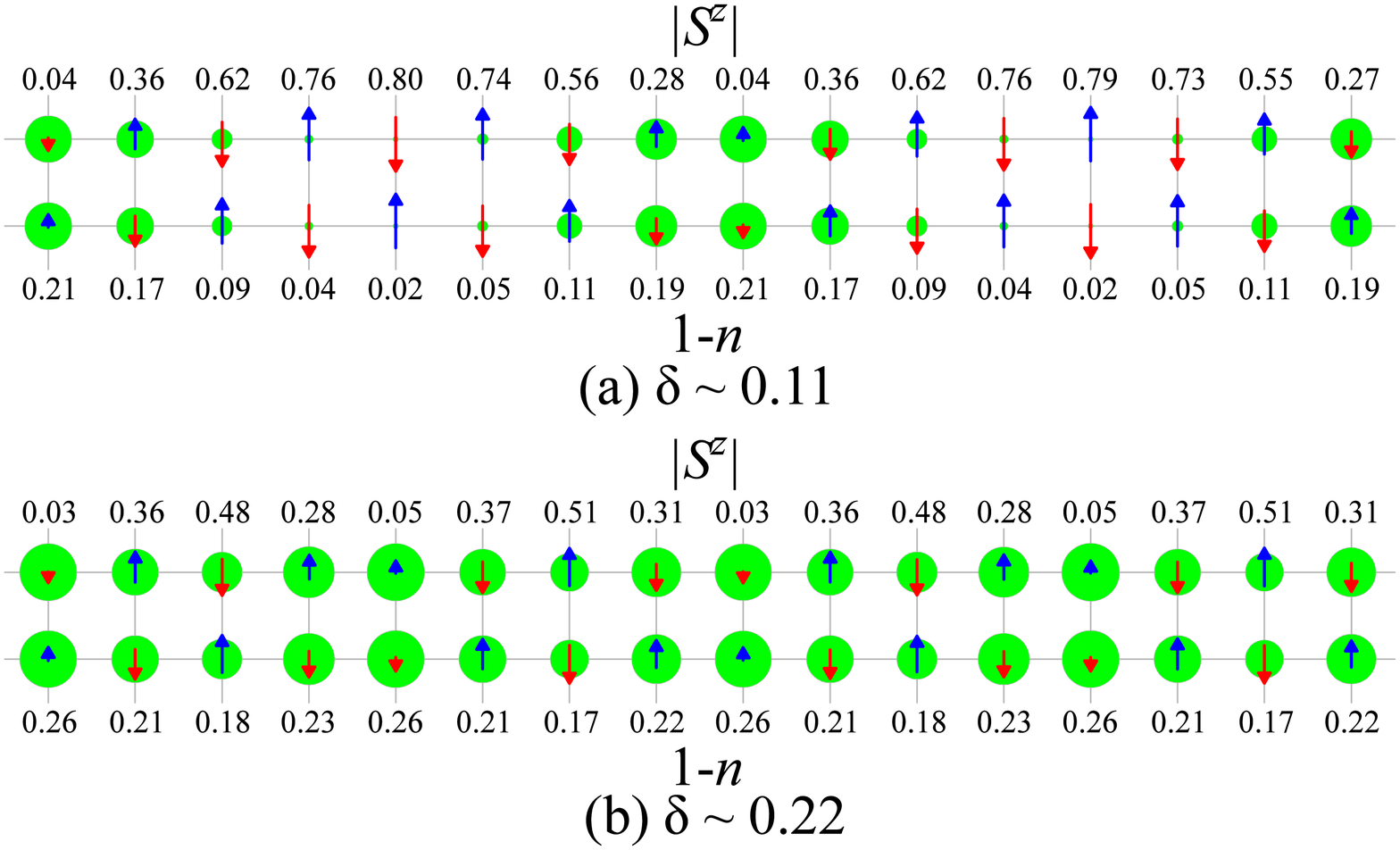} 
\caption{
(color online) 
Spin and charge orders at $\delta \sim 0.11 $ (upper panel) and $\delta \sim 0.22 $ (lower panel) on $16\times16$ lattice with $U=10$. 
The radius of every circle is proportional to the hole density $1-n$.
The length of every arrow is proportional to the spin density along $z$ direction $\left\vert S^{z}\right\vert $, and up and down arrows represent positive and negative $S^{z}$, respectively.  
The values of $\left\vert S^{z} \right\vert $ and $1-n$ are shown above and below the order plots, respectively. 
}
\label{fig:stripe} 
\end{figure*}

In order to show clear comparison, 
we pictorially depict the spin and charge orders at $\delta \sim 0.11 $ and $\delta \sim
0.22 $ in Fig. \ref{fig:stripe}.

\begin{figure}
\includegraphics[width=8cm]{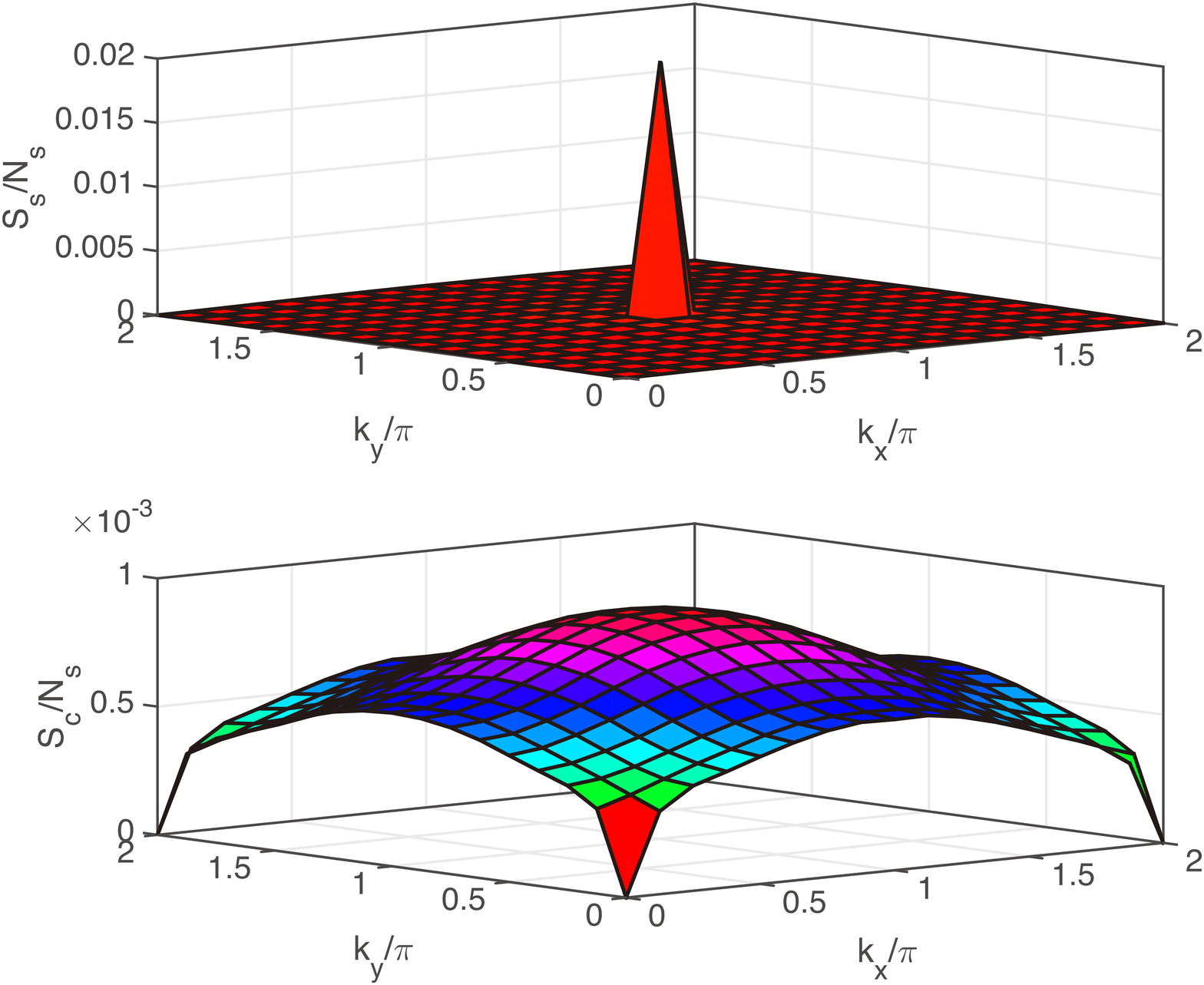} \caption{(color online) Spin (upper panel) and charge (lower panel) structure factors for the charge homogeneous superconducting state at $\delta\sim0.11$ on $16\times16$ lattice with $U=10$.
No prominent structure is visible in the charge correlations.}

\label{fig:Sc_homo} 
\end{figure}

To emphasize the difference between the charge inhomogeneous ground state, and the charge-homogeneous and superconducting excited state, we here show the charge and spin correlations of the charge-homogeneous excited state for an example at $\delta\sim 0.11$ and $U=10$ in Fig.~\ref{fig:Sc_homo}. 
The charge structure factor does not have an appreciable peak confirming the charge homogeneity, while the spin correlation has a peak at the commensurate wave number $(\pi,\pi )$. The coexistence of the superconductivity with the antiferromagnetic order around $\delta\sim 0.1$ was already found for the charge homogeneous state~\cite{Misawa2014PRB}.  
\section{\label{sec:summary}Summary and discussion}

We have proposed a method, which combines the tensor network and the
variational Monte Carlo method by taking advantage of the both to
study fermionic lattice models. In order to perform fast contraction
and preserve the lattice symmetry, we have introduced the FTTN into
the variational wave function. Our calculation shows that this combined
method substantially improves the accuracy in comparison to the accuracies
separately achieved by the conventional VMC calculation and the tensor
network.

Tensor network states usually satisfy entanglement area law which
may become inefficient to capture the large amount of entanglement
in itinerant fermionic systems. 
Recently, there exists an attempt to  alleviate the limitation of the area law by the PEPS simulations with the help of the energy extrapolation\cite{PEPSextrapolate} and it was applied to the issue of the competing phases in the doped Hubbard model\cite{stripeU8} 
The mVMC provides a more flexible reference
wave function instead of the basis of real space product states, so
that the combination with the mVMC extends the power of the tensor
network algorithms particularly for highly entangled correlated metals. Detailed comparisons with the single PEPS algorithm about the accuracy are left for future study.

We have applied the present method to study the ground state of the
hole doped Hubbard model. The ground states show coexisting stripe-type
charge and spin orders and weak $d$-wave superconducting order in the
lightly doped region. The results indicate that the strong superconducting
order realized by imposing spatial homogeneity becomes an excited
state and the true ground state is replaced by the state with the
superconducting correlation substantially weakened by the emergence
of the charge/spin stripe order with the period depending on the doping
concentration, which can be seen from Fig. \ref{fig:sc_r}. This compromised
ground state shows the universal feature of the strong-coupling superconductivity
that is subject to the spatial inhomogeneity including the phase separation
and charge/spin order. The high-$T_{{\rm c}}$ superconductivity in
the strongly correlated electron systems needs to overcome the simultaneous
charge inhomogeneity that weakens the superconductivity.
The excitation energy of the charge-uniform superconducting state is very small and roughly around 0.01 with small doping concentration dependence (see Fig.~\ref{fig:E_doping_ft}).   

Ref. \onlinecite{stripeU8} has studied Hubbard model at $U=8$ with
$\frac{1}{8}$ doping by various kinds of latest numerical methods,
which is consistent with the existence of stripe orders found here.
Throughout our calculation on the doped Hubbard model, we fix at $U=10$,
and we leave the $U$ dependence for future
work to further study the stability of the stripe order.

We have studied only the lattices with the sizes $2^n\times 2^n$.
Our calculation on $16\times 16$ lattice shows that the ground states below $\delta \sim 0.12$ and above $\delta \sim 0.2$ show the stripe orders with $(l_c, l_s)=(8,16)$ and $(l_c, l_s)=(4,8)$, respectively, while roughly in the region $0.12 <\delta<0.2$ the energy is convex implying the phase separation. However, one can speculate that other periodicities or structures of the stripes with the period between (8,16) and (4,8) that are not compatible with this size may have slightly lower energy filling the convexity
like $t$-$J$ model\cite{StripetJ} and precludes the phase separation in this region.
From the energy curve in Fig.~\ref{fig:E_doping_ft} the phase separation is expected as well in the region  $0<\delta < 0.12$, 
which looks more robust. 
Systematic studies along this line are left for future studies.
In addition, analyses on different lattice structures as well as effects of the intersite Coulomb interaction are intriguing future issues. In particular the intersite interaction may substantially change the behavior of the charge modulation.

\begin{acknowledgments}
We would like to thank Naoki Kawashima, Frank Pollmann and Ying-Jer Kao for stimulating discussion.
The authors thank the Supercomputer Center, the Institute for Solid
State Physics, the University of Tokyo for the facilities. This work
was financially supported by Japan Society for the Promotion of Science
through Program for Leading Graduate Schools (MERIT), the MEXT HPCI
Strategic Programs for Innovative Research (SPIRE), the Computational
Materials Science Initiative (CMSI) and Creation of New Functional
Devices and High-Performance Materials to Support Next-Generation
Industries (CDMSI). 
We thank the computational resources of the K
computer provided by the RIKEN Advanced Institute for Computational
Science through the HPCI System Research project (under the project
number hp130007, hp140215, hp150211, and hp160201). This work was
also supported by a Grant-in-Aid for Scientific Research (No. 22104010,
No. 22340090 and No. 16H06345) from Ministry of Education, Culture,
Sports, Science and Technology, Japan. 
\end{acknowledgments}

\appendix

\section{Tree tensor network state\label{Apdx:TTN}}

\begin{figure}
\includegraphics[width=8cm]{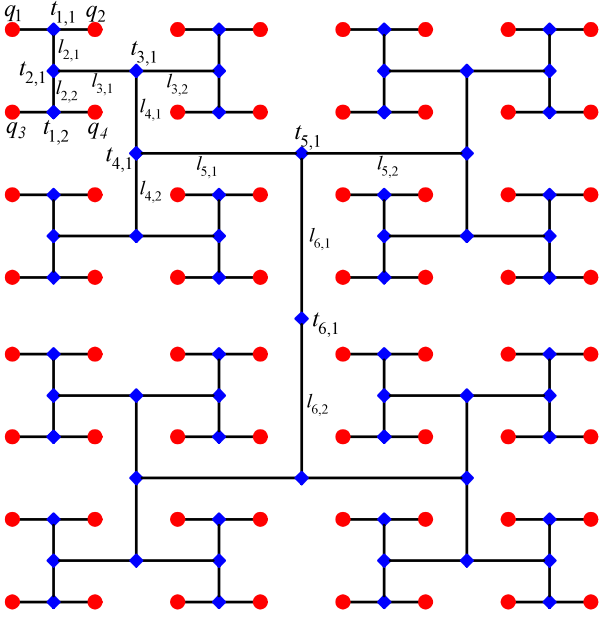} \caption{(color online) Example of a TTN for a $8\times8$ lattice. The red
solid circles represent the lattice sites, and the blue diamonds represent
the tensors in the TTN. $t_{1,i}$ is the leaf tensor which contains
2 physical indices, and $t_{2,i},t_{3,i},\ldots,t_{6,1}$ are internal-node
tensors which only contain virtual indices.}
\label{fig:TTN} 
\end{figure}

In this appendix, we will describe the tree tensor network state.
To reduce computational cost, we employ a binary tree in this work.
As an example, we consider a square lattice with $N_{s}=L\times L$ sites,
where each site has a local Hilbert space with dimension of $d$.
As shown in Fig. \ref{fig:TTN} for $8\times8$ lattice, the TTN is composed of a set of tensors 
$t_{j,i}$, where $j=1,2,\cdots,R\left(R=\log_{2}N_{s}\right)$, 
and $i=1,2,\ldots,N_{s}/2^{j}$. The TTN is connected as a binary
tree structure, which can be expressed as
\begin{eqnarray}
 &  & \mathcal{M}\left|q_{1},q_{2},\ldots,q_{N_{s}}\right\rangle \nonumber \\
 & = & \sum_{\left\{ l_{j,i}\right\} =1}^{D}\prod_{i=1}^{N_{s}/2}t_{1,i}\left(q_{2i-1},q_{2i},l_{2,i}\right)\nonumber \\
 &  & \left[\prod_{j=2}^{\left(R-1\right)}\prod_{i=1}^{N_{s}/2^{j}}t_{j,i}\left(l_{j,2i-1},l_{j,2i},l_{j+1,i}\right)\right]\nonumber \\
 &  & t_{R,1}\left(l_{R,1},l_{R,2}\right) \left|q_{1},q_{2},\ldots,q_{N_{s}}\right\rangle 
\end{eqnarray}
where $t_{1,i}$ is the leaf tensor which directly contains
2 physical indices (2 sites) and 1 virtual index, and 
$t_{2,i},t_{3,i},\ldots,t_{R,1}$
are internal-node tensors which only contain virtual indices.

The representation of a wave function with a TTN on real space basis
can be interpreted as real space coarse-graining transformation. Each
layer of node tensors reduces the effective lattice size by a factor
of 2, so the height (the number of the hierarchical levels) of the
tree is $\log_{2}\left(L\times L\right)$. Differently from coarse-graining
transformation, the node tensors in the TTN are not necessarily isometric.

The structure of the hierarchy of the node tensors are equivalent
to that of elimination tournament play. The total number of node tensors
in the binary tree is $N_{s}-1$.
Since there is no loop in the TTN, exact contraction is feasible.
If we employ the Monte Carlo sampling on the real space configuration,
the physical indices of the leaf tensors are fixed so every leaf tensor
becomes a vector. Therefore, we can start from the contraction of
the vector at the leaf tensor and then continue the contraction of
rank 3 tensors at the higher hierarchical levels, of which the computational
cost scales as $O\left(N_{s}D^{3}\right)$, where $D$ is the dimension
of the virtual indices. Since we only move one or two electrons in
every Monte Carlo step, we can reuse the intermediate result from
the tensor network contraction on the previous configuration. Therefore,
the computational cost can be reduced to $O\left(\log_{2}\left(N_{s}\right)D^{3}\right)$.

\section{Stochastic reconfiguration with conjugate gradient\label{Apdx:SRCG}}

To solve the SR equation Eq. (\ref{eq:SReq}) by a direct method,
one should calculate the overlap matrix $S_{kl}=\left\langle \Psi^{k}|\Psi^{l}\right\rangle $.
By introducing

\begin{equation}
\mathcal{O}_{k}\left(j\right)=\frac{\partial\left\langle x_{j}|\Psi\right\rangle /\partial\alpha_{k}}{\left\langle x_{j}|\Psi\right\rangle },
\end{equation}
$S_{kl}$ can be calculated as

\begin{eqnarray}
S_{kl} & = & \sum_{j}\rho\left(j\right)\mathcal{O}_{k}^{\dagger}\left(j\right)\mathcal{O}_{l}\left(j\right)\nonumber \\
 &  & -\left(\sum_{j=1}^{n_{s}}\rho\left(j\right)\mathcal{O}_{k}^{\dagger}\left(j\right)\right)\left(\sum_{m=1}^{n_{s}}\rho\left(m\right)\mathcal{O}_{l}\left(m\right)\right)\label{eq:Smatrix}
\end{eqnarray}
where $j$ is the sample index, and $\rho\left(j\right)$ is the weight
function for the importance sampling defined as

\begin{equation}
\rho\left(j\right)=\frac{\left\langle \Psi|x_{j}\right\rangle \left\langle x_{j}|\Psi\right\rangle }{\left\langle \Psi|\Psi\right\rangle }.
\end{equation}

The diagonal elements of $S$ matrix is the sample variance of $\mathcal{O}$

\begin{eqnarray}
        S_{kk} & = & \frac{1}{n_{s}}\sum_{j=1}^{n_{s}}\left(\mathcal{O}_{k}\left(j\right)-\left\langle \mathcal{O}_{k}\right\rangle \right)^{2},
\end{eqnarray}
and the variance of $S_{kk}$ is given by

\begin{eqnarray}
 &   & \text{Var} \left(S_{kk}\right) \nonumber \\
 & = & \frac{1}{n}(\left\langle \mathcal{O}_{k}^{4}\right\rangle -4\left\langle \mathcal{O}_{k}^{3}\right\rangle \left\langle \mathcal{O}_{k}\right\rangle +8\left\langle \mathcal{O}_{k}^{2}\right\rangle \left\langle \mathcal{O}_{k}\right\rangle ^{2}\nonumber \\
 &  & -4\left\langle \mathcal{O}_{k}\right\rangle ^{4}-\left\langle \mathcal{O}_{k}^{2}\right\rangle ^{2}).
\end{eqnarray}
If  $\text{Var}\left(S_{kk}\right)/S_{kk}$ is larger than a threshold, we
truncate the $k$-th parameter in the SR equation to stabilize
the optimization.

According to Eq. (\ref{eq:Smatrix}), the time cost for explicit construction
of $n_{p}\times n_{p}$ $S$ matrix is $O\left(n_{s}n_{p}^{2}\right)$,
and the time cost for solving the SR equation is $O\left(n_{p}^{3}\right)$,
while the memory cost is $O\left(n_{p}^{2}\right)$, where $n_{s}$
is the number of samples and $n_{p}$ is the number of variational
parameters. This is both the main time and memory consuming part,
when the number of variational parameters becomes large.

To reduce the cost, one can solve the SR equation iteratively by conjugate
gradient (CG) method, so that the explicit construction of $S_{kl}$ is
not required\citep{Neuscamman2012}. We only need to realize the matrix-vector
multiplication $\sum_{l=1}^{n_{p}}S_{kl}\gamma_{l}$, which can be
calculated by the Monte Carlo sampling as

\begin{eqnarray}
 &  & \sum_{l=1}^{n_{p}}S_{kl}\gamma_{l}\nonumber \\
 & = & \sum_{l=1}^{n_{p}}\sum_{j=1}^{n_{s}}\rho\left(j\right)\mathcal{O}_{k}^{\dagger}\left(j\right)\mathcal{O}_{l}\left(j\right)\gamma_{l}\nonumber \\
 & - & \sum_{l=1}^{n_{p}}\sum_{j=1}^{n_{s}}\rho\left(j\right)\mathcal{O}_{k}^{\dagger}\left(j\right)\left(\sum_{m=1}^{n_{s}}\rho\left(m\right)\mathcal{O}_{l}\left(m\right)\right)\gamma_{l}.\label{eq:mat-vec-prod}
\end{eqnarray}
To reduce the memory cost, one can change the summation order of the
variational parameters and samples as

\begin{eqnarray}
 &  & \sum_{l=1}^{n_{p}}S_{kl}\gamma_{l}\nonumber \\
 & = & \sum_{j=1}^{n_{s}}\rho\left(j\right)\mathcal{O}_{k}^{\dagger}\left(j\right)\sum_{l=1}^{n_{p}}\mathcal{O}_{l}\left(j\right)\gamma_{l}\nonumber \\
 & - & \sum_{j=1}^{n_{s}}\rho\left(j\right)\mathcal{O}_{k}^{\dagger}\left(j\right)\sum_{l=1}^{n_{p}}\left(\sum_{m=1}^{n_{s}}\rho\left(m\right)\mathcal{O}_{l}\left(m\right)\right)\gamma_{l}.\label{eq:mat-vec-prod-reorder-1}
\end{eqnarray}
As a result, the computational cost is reduced from $O\left(n_{s}n_{p}^{2}+n_{p}^{3}\right)$
to $O\left(n_{s}n_{p}n_{\rm iter}\right)$, and the memory cost is reduced from
$O\left(n_{s}n_{p}+n_{p}^{2}\right)$ to $O\left(n_{s}n_{p}\right)$.
By this application of the CG method, it allows an
efficient way of solving Eq. (\ref{eq:SReq}).

\section{Fast update of Pfaffian\label{Apdx:recalPf}}

In each Monte Carlo sample of real space configuration 
\begin{equation}
\left|x\right\rangle =c_{r_{1}\sigma_{1}}^{\dagger}c_{r_{2}\sigma_{2}}^{\dagger}\cdots c_{r_{N_{e}}\sigma_{N_{e}}}^{\dagger}\left|0\right\rangle ,
\end{equation}
one can calculate $\left\langle x|\phi_{{\rm pair}}\right\rangle $
by the computation of Pfaffian of the matrix

\begin{equation}
A_{ij}=f_{r_{i}\sigma_{i},r_{j}\sigma_{j}}-f_{r_{j}\sigma_{j},r_{i}\sigma_{i}}.
\end{equation}
If a new sample $x'$ is proposed by changing the position of one
electron in $x$, then one can calculate ${\rm Pf}\left(B\right)=\left\langle x|\phi_{{\rm pair}}\right\rangle $
by

\begin{equation}
{\rm Pf}\left(B\right)={\rm Pf}\left(A\right)\sum_{m}A_{\alpha m}^{-1}B_{\alpha m},\label{eq:Pfupdate}
\end{equation}
where matrices $A$ and $B$ have same elements except for $\alpha$-th
row and column. From Eq. (\ref{eq:Pfupdate}), one can update Pfaffian
at the cost of $O\left(M\right)$, where $M$ is the dimension of
the matrix.

This fast update of Pfaffian is only possible if the inverse of $A$
is known, but the inverse need only to be directly computed once,
as it can be efficiently updated when one row or one column of the
matrix changes by using the Sherman-Morrison formula\cite{ShermanMorrison1949},

\begin{equation}
\left(A+uv^{T}\right)^{-1}=A^{-1}-\frac{A^{-1}uv^{T}A^{-1}}{1+v^{T}A^{-1}u},\label{SermanMorrison}
\end{equation}
where $u$ and $v$ are column vectors, where the computational cost
of Eq. (\ref{SermanMorrison}) is $O\left(M^{2}\right)$. 
The detailed explanation of the update technique can be found in Ref. \onlinecite{SMorita2015}
However, during the update of matrix $A$, the round-off error accumulates.
As a result, the computation of matrix inverse would be inaccurate,
so that the fast update of Pfaffian would contain large error.

To solve this problem, we estimate the condition number by matrix
1-norm

\begin{equation}
\kappa\left(A\right)=\left\Vert A^{-1}\right\Vert _{1}\left\Vert A\right\Vert _{1},
\end{equation}
where the 1-norm of matrix is defined as 
\[
\left\Vert A\right\Vert _{1}=\underset{1\leq j\leq n}{{\rm max}}\sum_{i=1}^{m}\left|a_{ij}\right|.
\]

If the condition number becomes larger than threshold $\kappa\left(A\right)>\kappa_{c}$,
the Pfaffian of $A$ should be computed from scratch, instead of using
fast update procedure. The computational cost of 1-norm scales as
$O\left(M^{2}\right)$, which is the same as the computation cost
of the matrix inverse by the Sherman-Morrison formula, so that the
estimate of the condition number does not increase the order of the
cost in the update of the Pfaffian.

\section{Rescaling of the variational parameters \label{Apdx:rescaling}}

If the condition number of overlap matrix $S_{kl}=\left\langle \Psi^{k}|\Psi^{l}\right\rangle $
is large, the solution of the SR equation would be inaccurate. Therefore,
to stabilize the optimization, we shift the diagonal elements in $S$

\begin{equation}
S\rightarrow S+\epsilon I,\label{eq:diagShift}
\end{equation}
where $\epsilon$ is a small constant.

For a real space sample $x$, the derivative of every tensor can be
expressed as

\begin{eqnarray}
D_{\mu}^{i} & = & \frac{\partial\left\langle x|\Psi\right\rangle /\partial t_{\mu}^{i}}{\left\langle x|\Psi\right\rangle }\nonumber \\
 & = & \frac{{\rm Tr}'\left(\prod_{j\neq i}t_{\nu}^{j}\left(x\right)\right)}{{\rm Tr}\left(\prod_{j}t_{\nu}^{j}\left(x\right)\right)},\label{eq:tensorDiff}
\end{eqnarray}
where $t_{\mu}^{i}$ is the $\mu$-th element of the $i$-th tensor.
The ${\rm Tr}$ in the denominator is to trace out virtual indices
of all tensors, and the ${\rm Tr}'$ in the numerator is to trace out virtual
indices of all tensors except $t^{i}$, and $\mu\left(\nu\right)$
is the simplified notation of all virtual indices of one tensor. If
$t^{i}$ is rescaled by a factor of $\eta^{i}$

\begin{equation}
t^{i}\rightarrow\eta^{i}t^{i},
\end{equation}
the wave function is unchanged, but the derivative of the corresponding
tensor has been rescaled as

\begin{equation}
D^{i}\rightarrow D^{i}/\eta^{i}.
\end{equation}
If inappropriate rescaling factor $\eta^{i}$ is employed, the derivative
of every tensor could be significantly different in the order of magnitude.
Therefore, the shift of diagonal elements in $S$ as Eq. (\ref{eq:diagShift})
will suppress the change of parameters in tensors which have small
derivatives.

To change parameters efficiently while stabilizing the optimization
in the SR procedure, we determine the rescaling factor according to
the amplitude of the derivative of each tensor as

\begin{equation}
\eta^{i}=\underset{\mu}{{\rm max}}\left(D_{\mu}^{i}\right),
\end{equation}
instead of rescaling the parameters according to the amplitude of
tensor elements as

\begin{equation}
\eta^{i}=1/\underset{\mu}{{\rm max}}\left(t_{\mu}^{i}\right).
\end{equation}
As a result, the derivative of every tensor would be more or less
in the same order of magnitude, so that SR method can optimize elements
of every tensor efficiently.

\end{document}